\documentclass[preprint,aps,12pt,showpacs,nofootinbib,tightenlines]{revtex4}
\usepackage{graphicx}

\def \etal{{\it et al.}}

\newcommand{\TeV}{{\rm TeV}}
\newcommand{\GeV}{{\rm GeV}}
\begin{document}

\title{Neutral Higgs production on LHC in the two-Higgs-doublet model with spontaneous $CP$ violation}
\author{Shou-Shan Bao}
\author{Yue-Liang Wu} \email{ylwu@itp.ac.cn}
\affiliation{ Kavli Institute for Theoretical Physics China (KITPC)
\\ Key Laboratory of Frontiers in Theoretical Physics \\ Institute of
Theoretical Physics, Chinese Academy of Science, Beijing,100190,
P.R.China}
\date{\today}
\begin{abstract}
Spontaneous CP violation motivates the introduction of two Higgs
doublets in the electroweak theory, such a simple extension of the
standard model has five physical Higgs bosons and rich CP-violating
sources. Exploration on more than one Higgs boson is a direct
evidence for new physics beyond the standard model. The neutral
Higgs production at LHC is investigated in such a general two Higgs
doublet model with spontaneous CP violation, it is shown that the
production cross section and decays of the neutral Higgs boson can
significantly be different from the predictions from the standard
model.
\end{abstract}
 \maketitle

\section{Introduction}

In the standard model(SM), the fermions and gauge bosons get masses
through Higgs mechanism with a single weak-isospin doublet Higgs
field. After the electroweak symmetry breaking, three Goldstone
modes were absorbed to build up the longitudinal W and Z gauge
bosons, only one physical scalar called the SM Higgs boson is left.
Although the value of the Higgs mass can not be predicted in the SM,
for the theoretical self-consistence, the unitarity\cite{unitarity}
require $m_h<1\, \TeV$. On the other hand, the analysis of all the LEP
measurements\cite{lep} leads to the best fitting Higgs mass
$m_h=114^{+69}_{-45}\GeV$ or the one-side $95\%$ CL upper limit of
$m_h<260\GeV$. Once the Higgs mass is known in the SM, the
properties of the Higgs boson, such as the decay width and
production cross section, can be predicted. Nevertheless, if there
exists new physics beyond the SM, the production cross section and
decay width of the Higgs boson as well as the mass constraint to the
Higgs boson can be different.

It has been shown that if the SM Higgs mass lies between 130 and
200Gev\cite{higgsarg}, the SM can in general be valid at energy
scales all the way up to the Planck scale. Nevertheless, the SM
cannot be a fundamental theory, there are still some unknown puzzles
in the SM, such as the origin of $CP$ violation, the smallness of
neutrino masses, the dark matter and so on. They all suggest the
existence of new physics beyond the SM. Thus, many extensions or
modifications of the SM have been studied. In this paper, we are
going to focus on the simplest extension of the SM with adding an
extra Higgs double motivated from spontaneous $CP$
violation(SCPV)\cite{TDLEE,SW,YLW1,YLW2,cpv}, such a general two
Higgs doublet model (2HDM) with spontaneous $CP$ violation is also
called type III 2HDM. It has been shown that if one Higgs doublet is
needed for the mass generation, then the additional extra Higgs
doublet is necessary for the origin of $CP$ violation, so that the
$CP$ violation is originated from a single relative phase of two
vacuum expectation values, which gives not only an explanation for
the Kobayashi-Maskawa $CP$-violating mechanism\cite{KM} in the
standard model, but also leads to a new type of $CP$-violating
source\cite{YLW1,YLW2} which has been studied broadly.

The complex Higgs doublets in the type III 2HDM are generally
expressed as~\cite{YLW1,YLW2,PR}
\begin{equation}
\Phi_1=\left(\begin{array}{c} \phi_1^+\\ \phi_1^0
\end{array}\right), \, \, \, \Phi_2=\left(\begin{array}{c} \phi_2^+\\
\phi_2^0\end{array}\right),
\end{equation}
and the Higgs potential is
\begin{eqnarray}
V&=&-\mu _1 ^2\Phi_1^\dag\Phi_1-\mu _2^2\Phi_2^\dag\Phi_2
-\frac{\mu_{12}^2}{2} \Phi_1^\dag\Phi_2 - \frac{\mu_{12}^{\ast
2}}{2}\Phi_2^\dag\Phi_1
+{{\lambda }_1}{\left(\Phi_1^\dag\Phi_1\right)^2}+ {{\lambda }_2}{\left(\Phi_2^\dag\Phi_2\right)^2}\nonumber\\
&&+\frac{{\lambda
}_3}{4}{(\Phi_1^\dag\Phi_2+\Phi_2^\dag\Phi_1)^2}-\frac{{\lambda
}_4}{4}{(\Phi_1^\dag\Phi_2-\Phi_2^\dag\Phi_1)^2}+{{\lambda
}_5}\Phi_1^\dag\Phi_1\Phi_2^\dag\Phi_2\nonumber\\
&&+\frac{1}{2}(\lambda_6\Phi_1^\dag\Phi_1+\lambda_7\Phi_2^\dag\Phi_2)(\Phi_1^\dag\Phi_2+\Phi_2^\dag\Phi_1).
\end{eqnarray}
The Yukawa interaction terms have the following general form
\begin{eqnarray}
\mathcal{L}_{Y}
&=&\eta_{ij}^{(k)}\bar{\psi}_{i,L}\tilde{\Phi}_kU_{j,R}+\xi_{ij}^{(k)}\bar{\psi}_{i,L}\Phi_kD_{j,R}+H.c.,\label{eq:oyukawa}
\end{eqnarray}
where $\eta_{ij}^{(k)}$ and $\xi_{ij}^{(k)}$ are real Yukawa
coupling constants, so that the interactions are $CP$ invariant. The
major issue with respect to the model is that it allows flavor
changing neutral current (FCNC) at the tree level through the
neutral Higgs boson exchanges, which should strongly be suppressed
based on the experimental observations. In order to prevent the FCNC
at the tree level, an {\it ad hoc} discrete symmetry\cite{discrete}
is often imposed:
\begin{eqnarray}
  \Phi_1\to-\Phi_1 &\mbox{and}& \Phi_2\to \Phi_2,\nonumber\\
  U_R\to -U_R &\mbox{and}& D_R\to\mp D_R,
\end{eqnarray}
which leads to the so called Type I and Type II 2HDM relying on
whether the up- and down-type quarks are coupled to the same or
different Higgs doublet respectively. Some interesting phenomena for
various cases in such types of models without FCNC have been
investigated in detail in Refs.\cite{barger1, barger2}. When the
discrete symmetry was introduced, there will be $\mu_{12}=0$ and
$\lambda_6=\lambda_7=0$ which implies no spontaneous CP violation
any more \cite{noscpv}. It should be noted that the supersymmetry
also requires more than one Higgs doublet. The Higgs sector and the
relevant Yukawa interactions in the minimal supersymmetric standard
model (MSSM) is analogous to the type II 2HDM. As the FCNC is the
interesting phenomena observed in experiments in the weak
interactions though it is strongly suppressed, we shall abandon the
discrete symmetries and consider the small off-diagonal Yukawa
couplings concerning the FCNC, the naturalness for such small Yukawa
couplings may be understood from the approximate global U(1) family
symmetries\cite{YLW1,YLW2,YLW3,FS,HW}. This may be explained as
follows: if all the up-type quarks and also the down-type quarks
have the same masses and no mixing, the theory has an U(3) family
symmetry for three generation, while when all quarks have different
masses but remain no mixing, the theory has the $U(1) \otimes U(1)
\otimes U(1)$ family symmetries and the Cabibbo-Kobayashi-Maskawa
quark-mixing matrix is a unit matrix, in this case both the direct
FCNC and induced FCNC are absent. In the real world, there are some
FCNC processes observed, thus the U(1) family symmetries should be
broken down. As all the observed FCNC processes are strongly
suppressed, the theory should possess approximate U(1) family
symmetries with small off-diagnoal mixing among the generations. In
this sense, the approximate U(1) family symmetries are enough to
ensure the naturalness of the observed smallness of FCNC.

After spontaneous symmetry breaking, the neutral Higgs bosons will
get the vacuum expectation values
\begin{equation}
\langle\phi_1^0\rangle=\frac{1}{\sqrt{2}} v_1 e^{i \delta_1}, \, \,
\, \langle\phi_2^0\rangle= \frac{1}{\sqrt{2}} v_2 e^{i \delta_2},
\end{equation}
where one of the phases can be rotated away due to the global U(1)
symmetry. Without losing generality, we may take $\delta_1=0$ and
$\delta_2=\delta$. It is then convenient to make a unitary
transformation
\begin{equation}
\label{eq:higgs0}
\left(\begin{array}{c}
H_1\\
H_2
\end{array}\right)=U\left(\begin{array}{c}
\Phi_1\\
\Phi_2
\end{array}\right),\label{eq:transform}
\end{equation}
with
\begin{equation}
 U= \left(
\begin{array}{cc}
\cos\beta&\sin\beta e^{-i\delta}\\
-\sin\beta &\cos\beta e^{-i\delta}
\end{array}
\right)
\end{equation}
and $\tan \beta =v_2/v_1$. After making the above transformation, we
can re-express the Higgs doublets as follows:
\begin{equation}
\label{eq:higgs}
H_1=\frac{1}{\sqrt{2}}\left( \begin{array}{c} 0\\ v+\phi_1^0
\end{array}\right)+\mathcal{G}, \, \, \, H_2=\frac{1}{\sqrt{2}}\left(
\begin{array}{c} \sqrt{2} H ^+\\ \phi_2^0+i \phi_3^0
\end{array}\right),
\end{equation}
with $v^2=v_1^2+v_2^2$ and $v\simeq 246\GeV$ which is the same as in
the standard model. Thus the Higgs doublet $H_1$ in the new basis
plays the role of the standard model Higgs and gives masses to the
gauge bosons ($m_W= g v/2$) and quarks and leptons. The Fermi
constant is then given by the same value as in the standard model
$G_F=g^2/(4\sqrt{2}m_W^2) =1/(\sqrt{2}v^2)$. The Higgs field
$\mathcal{G}$ are the goldstone particles absorbed by the gauge
bosons, while $H^{\pm}$ are mass eigenstates of the charged scalar
Higgs, and ($\phi_1^0, \phi_2^0, \phi_3^0$) are the neutral Higgs
bosons in the electroweak eigenstates, they are in general not the
same ones ($h, H, A$) in the mass eigenstates, but can be expressed
as linear combinations of the mass eigenstates ($h, H, A$) via an
orthogonal SO(3) transformation which depends on the $\lambda_i$s
and $\mu_i^2$s in the Higgs potential. In the new basis after the
unitary transformation, the phase $\delta$ appears in the Yukawa
coupling terms
\begin{eqnarray}
\mathcal{L}_{Y}  &=&\eta_{ij}^U\bar{\psi}_{i,L}\tilde{H}_1U_{j,R}+\eta_{ij}^D\bar{\psi}_{i,L}H_1D_{j,R}
+\xi_{ij}^U\bar{\psi}_{i,L}\tilde{H}_2U_{j,R}\nonumber\\
  &&+\xi_{ij}^D\bar{\psi}_{iL}H_2D_{j,R}+H.c.,
\end{eqnarray}
where
\begin{eqnarray}
  \eta_{ij}^U&= & \eta_{ij}^{(1)}\cos\beta+\eta_{ij}^{(2)}e^{-i\delta}\sin\beta \equiv \sqrt{2}M_{ij}^U/v ,\nonumber \\
  \xi_{ij}^U&= &-\eta_{ij}^{(1)} e^{-i\delta}\sin\beta+\eta_{ij}^{(2)}\cos\beta,\nonumber \\
  \eta_{ij}^D&= & \xi_{ij}^{(1)}\cos\beta+\xi_{ij}^{(2)} e^{-i\delta}\sin\beta \equiv \sqrt{2}M_{ij}^D/v, \nonumber\\
  \xi_{ij}^D&= &-\xi_{ij}^{(1)} e^{-i\delta}\sin\beta+\xi_{ij}^{(2)}\cos\beta.
\end{eqnarray}
As the Yukawa coupling terms $\eta^{U}$ and $\eta^{D}$ become
complex due to the vacuum phase $\delta$, the resulting mass
matrices are also complex. It then requires a unitary transformation
to diagonalize the mass matrices for transforming the quark and
lepton fields from the weak interaction states to the mass
eigenstates. In the mass eigenstates, the Yukawa interaction terms
are given by
\begin{eqnarray}
\label{Yukawa}
  \mathcal{L}_{Y}&=&\bar{U}_{L}\frac{m^U}{v}U_{R}(v+
  \phi_1^0)+\bar{D}_{L}\frac{m^D}{v}D_{R}(v+\phi_1^0)\nonumber\\
  &&+\frac{1}{\sqrt{2}}\bar{U}_{L} {\xi}^U U_{R}(\phi_2^0-i \phi_3^0)+\bar{D}_{L}
  \hat{\xi}^U U_{R}H^-\nonumber\\
  &&+\bar{U}_{L}{\hat{\xi}^D} D_{R}H^++\frac{1}{\sqrt{2}}\bar{D}_{L} {\xi}^DD_{R}
  (\phi_2^0+i \phi_3^0)+H.c.,
  \label{eq:lagrangian}
\end{eqnarray}
with
\begin{eqnarray}
\hat{\xi}^{U}={\xi}^{U} V_{\mbox{CKM}},\qquad \hat{\xi}^D =
V_{\mbox{CKM}}{\xi}^{D},\label{eq:yukawa}
\end{eqnarray}
and
\begin{eqnarray}
  U_{L(R)}=\left(u_{L(R)}, c_{L(R)}, t_{L(R)}\right), \qquad D_{L(R)}=\left(d_{L(R)}, s_{L(R)}, b_{L(R)}\right).
\end{eqnarray}
it can be seen that the scalar $\phi_1^0$ plays the role of the
Higgs in the SM except considering the large mixing effects among
$\phi_1^0$, $\phi_2^0$ and $\phi_3^0$, which will be discussed later
on. Here we shall use $\xi^{U(D)}$ and the masses of quarks as the
independent input parameters instead of  the original Yukawa
couplings in Eq.(\ref{eq:oyukawa}) and the parameter $\beta$.

Note that the Type II 2HDM can be regarded as a special case of the
Type III 2HDM with spontaneous $CP$ violation by setting
$\eta_{ij}^{(1)} =0$ and $\xi_{ij}^{(2)}=0$ which are ensured by a
discrete symmetry,
\begin{eqnarray}
  \eta_{ij}^U&= & \eta_{ij}^{(2)} \sin\beta \equiv \sqrt{2}M_{ij}^U/v,\quad
  \xi_{ij}^U = \eta_{ij}^{(2)}\cos\beta \equiv \sqrt{2}M_{ij}^U/v \cot\beta ,\nonumber \\
  \eta_{ij}^D&= & \xi_{ij}^{(1)}\cos\beta \equiv \sqrt{2}M_{ij}^D/v, \quad
  \xi_{ij}^D = -\xi_{ij}^{(1)} \sin\beta \equiv -\sqrt{2}M_{ij}^D/v \tan\beta.
\end{eqnarray}
with $M_{ij}^{U}$ and $M_{ij}^D$ being the mass matrices. Thus in
the Type II 2HDM, the Yukawa couplings are almost fixed by the
masses of quarks and Cabibbo-Kobayashi-Maskawa matrix elements, the
angle $\beta$ or the ratio of two vacuum expectation values is
manifestly an important parameter as it uniquely characterizes the
amplitude of Yukawa couplings for the up-type quarks $\xi_{ij}^U$
and down-type quarks $\xi_{ij}^D$ in the mass eigenstates. The
similar situation occurs for the minimal supersymmetric standard
model(MSSM). When parameterizing the Yukawa couplings by using the
quark mass scales,
\begin{eqnarray}
\xi_{ij}^{U,D}\equiv \lambda_{ij} \sqrt{2m_i m_j }/v,
\end{eqnarray}
where the smallness off-diagonal elements is characterized by the
hierarchical mass scales of quarks and the parameters
$\lambda_{ij}$. In terms of this parametrization, one has the
following simple relations in the Type II 2HDM
\begin{eqnarray}
\xi^t \equiv \xi_{tt}^U&= & \sqrt{2}m_t/v \cot\beta,\quad \lambda_t \equiv \lambda_{tt}^U \sim \cot\beta \nonumber \\
\xi^b \equiv \xi_{bb}^D&= & \sqrt{2}m_b/v \tan \beta,\quad \lambda_b
\equiv \lambda_{bb}^D \sim -\tan\beta.
\end{eqnarray}
The situation is obviously different from the general Type III 2HDM
with spontaneous $CP$ violation, where each Higgs doublet couples to
both up-type and down-type quarks, there are physically meaningful
Yukawa coupling constants which are twice as the ones in the Type II
2HDM, thus the dependence on parameter $\beta$ is not manifest as in
the Type II 2HDM. There are more free parameters in the Yukawa
interactions, they are in general determined only through various
experiments like in the standard model. Whereas it provides, from
the phenomenological points of view, an interesting window for
exploring possible new physics effects inspired from the Type III
2HDM with spontaneous $CP$ violation. It is seen that an
additionally imposed discrete symmetry is much stronger than the
$CP$ symmetry in the 2HDM. It would be interesting to have a
detailed study for both the type II (or type I) 2HDM and type III
2HDM, as the type II or type I model was motivated from the
assumption of natural flavor conservation, while type III model was
initiated from the origin of $CP$ violation with spontaneous
symmetry breaking.

It was observed in the Type III 2HDM \cite{YLW1,YLW2,YLW3} that the
charged Higgs interactions involving the Yukawa couplings
$\hat{\xi}^{U(D)}$ in Eq.(\ref{eq:lagrangian}) lead to a new type of
$CP$-violating FCNC even if the neutral current couplings
$\xi^{U(D)}$ are diagonal. For the parameters concerning the third
generation, we may express as
\begin{eqnarray}
\xi^{t}&\equiv & \xi^{U}_{tt}, \quad \xi^{t} = |\xi^{t}|e^{\delta_t}\, ,  \nonumber \\
\xi^{b}&\equiv & \xi^{D}_{tt}, \quad \xi^{b} =
|\xi^{b}|e^{\delta_b}.
\end{eqnarray}
The general constraints on the FCNC and the relevant parameter
spaces have been investigated in
\cite{PR,BCK,WUZHOU,2hdmeff,YLW4,b2gamma}. Here we may consider the
following three typical parameter spaces for the neutral Yukawa
couplings of $b$-quark and $t$-quark
${\xi^q}/{\sqrt{2}}=\lambda_{q}m_q/v$,
\begin{eqnarray}
\mbox{Case\quad A}:\quad |\xi^{t}/{\sqrt{2}}|&=& 0.2 (\lambda_t=0.3);\quad \ \,\quad |\xi^{b}/{\sqrt{2}}| =0.5 (\lambda_b=30),\nonumber\\
\mbox{Case\quad B}:\quad |\xi^{t}/{\sqrt{2}}|&=&0.1 (\lambda_t=0.15);\quad\quad |\xi^{b}/{\sqrt{2}}|=0.8 (\lambda_b=50), \label{eq:case}\\
\mbox{Case\quad C}:\quad
|\xi^{t}/{\sqrt{2}}|&=&0.01 (\lambda_t=0.015); \quad
|\xi^{b}/{\sqrt{2}}|=1.0 (\lambda_b=60),\nonumber
\end{eqnarray}
 which is consistent with the current experimental constraints in
the flavor sector including the $B$ meson decays\cite{B2PV,B2VV}
even when the neutral Higgs masses are taken to be the typical low
values
\begin{eqnarray}
  m_{A}=120\GeV,\quad m_h=115\GeV,\quad m_{H}=160\GeV.
\end{eqnarray}

 For the charged Higgs
mass, when $\frac{1}{\sqrt{2}}|\xi^{t}|\simeq 0.2$(or
$|\lambda_{t}|\sim 0.3$), the lower bound for the charged Higgs mass
can reach to be about $m_{H^+} \sim 160$ GeV from the
$B^0-\bar{B}^0$ mixing at the $1\sigma$ level (or about
$m_{H^+}\sim60$GeV at $2\sigma$ level). In general, a smaller value
of $|\xi^{t}|$ leads to a lower bound on the charged Higgs mass. The
strong constraints may arise from the radiative bottom quark decay
$b\to s\gamma$. In fact, its mass was found to be severely
constrained from the $b\to s\gamma$ decay in the Type II 2HDM, the
lower bound on the charged Higgs mass can be as large as $m_{H^+}
\simeq 350$ GeV, which is corresponding to the special case in the
Type III 2HDM with the parameter $|\xi^t||\xi^b| \sim 0.02$(or
$|\lambda_t\lambda_b|\sim1$) and a relative phase $\delta_t-\delta_b
= 180^{\circ}$. Nevertheless, the constraints can significantly be
relaxed in the Type III 2HDM due to the freedom of the parameters
$\xi^t$ and $\xi^b$ as well as their relative phase
$(\delta_t-\delta_b)$. In general, when the combined parameter
$|\xi^t||\xi^b|$ becomes smaller, the resulting bound to the charged
Higgs mass goes to be lower. While an interesting feature arises in
the Type III 2HDM, when the relative phase makes the charged Higgs
amplitude to interfere destructively with the standard model
amplitude, the allowed charged Higgs mass can remain small even for
a large value of combined parameter $|\xi^t||\xi^b|$. For instance,
when $|\xi^t||\xi^b| \lesssim 0.025$(or
$|\lambda_t\lambda_b|\lesssim1.0$), the allowed charged Higgs mass
can be in all range for a large range of the relative phase
($\delta_t-\delta_b \simeq \pi/4 \sim \pi/2$)\cite{BCK}, even when
taking the combined parameter to be large $|\xi^t\xi^b| \simeq
0.07$(or $|\lambda_t\lambda_b|\sim3$), the resulting bound on the
charge Higgs mass can still be as low as $m_{H^+} \sim 100$ GeV for
a certain range of the relative phase. Some strong constraints to
the charged Higgs mass may arise from the neutron electric dipole
momentum, but we shall not consider such possible constraints as it
involves large uncertainties caused by the hadronic matrix elements
and also receives various contributions from several CP-violating
sources in the Type III 2HDM\cite{YLW1}. Some upper limit on the
charged Higgs mass may arise from the $\rho$-parameter\cite{BCK},
which needs a more precise measurement.

In 2HDM, there are three neutral and one charged  Higgs bosons. The
charged Higgs is totally different to the particles in SM, and its
effect to lower-energy phenomenology and direct search have been
studied by many authors. As there are more neutral Higgs, in paper
\cite{paire_pro} the authors discussed the pair production of the
neutral Higgs $gg\to hh$ which is sensitive to the triple couplings
in the Higgs potential. In our present paper, as we only study the
neutral Higgs production and decays, which does not involve the
charged Higgs boson and the triple couplings in the potential at
lowest order, thus we may consider the allowed parameter space of
$\xi^t$ and $\xi^b$ to be as large as possible, and take the
combined parameter $|\xi^t||\xi^b|$ to range from $|\xi^t||\xi^b|
=0.02$ to $|\xi^t||\xi^b| = 0.2$, which is covered from the above
given three typical parameter spaces. Where the values of $|\xi^t|$
are taken to be small so as to fit the constraints from the
$B^0-\bar{B}^0$ mixing. For a large value of $|\xi^b|$, it may
naturally be resulted for a large value of $\tan\beta\sim 30$, but
it is not a necessary requirement in the Type III 2HDM. For any
given values of $\tan\beta$, one can always find appropriate
parameter space of $\xi^{(1)}_{tt}$ and $\xi^{(2)}_{tt}$, so as to
fit the bottom quark mass $m_b = \sqrt{2}|\eta_{bb}^D|/v$ and
meanwhile allow a large Yukawa coupling $|\xi^b|\equiv
|\xi_{bb}^D|$. For the same reason, one can find the appropriate
parameter space of $\eta^{(1)}_{tt}$ and $\eta^{(2)}_{tt}$ to yield
a small Yukawa coupling $|\xi^t|\equiv |\xi_{tt}^U|$ with
simultaneously fitting the top quark mass. Therefore, we shall take
the independent parameters $|\xi^t|$ and $|\xi^b|$ as the free
parameters in the Type III 2HDM instead of using the parameter
$\tan\beta$ which is unique in the Type II 2HDM and MSSM.

In the following sections, we shall calculate the neutral Higgs
productions and decays with the above three typical Yukawa couplings
and free neutral Higgs Masses ($<1$TeV). We first consider the Higgs
production in section \ref{section:productions}, and then the Higgs
decay in section \ref{section:decays}. In section
\ref{section:mixings} we shall discuss the effects of the mixing
between the neutral Higgs bosons. Our conclusion is presented in the
final section.

\section{The Higgs Productions\label{section:productions}}

\begin{figure}[htb]
\includegraphics[scale=0.5]{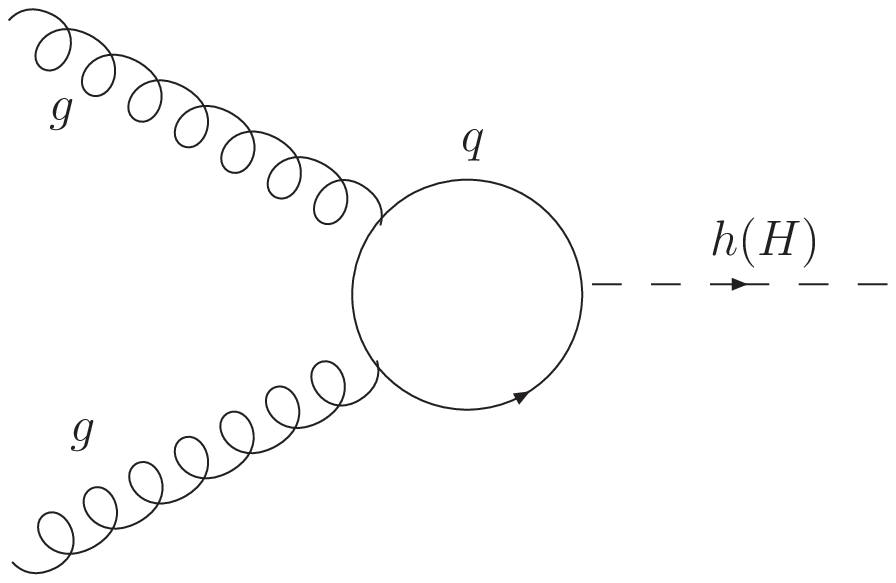}
\caption{The Feynman diagram of the Higgs production with
Gluon-Gluon Fusion.}
  \label{fig:gg2h}
\end{figure}

According to QCD, the quarks and gluons are the fundamental degrees
of freedom to participate in strong interactions at high energy, the
QCD parton model plays a pivotal role in understanding hadron
collisions\cite{Han:2005mu,collider}. Due to the gluon luminosity,
the gluon fusion is the main production channel of Higgs bosons in
proton-proton collisions throughout the entire Higgs mass range both
in SM and 2HDM, and the first prediction for the production cross
section of the SM Higgs was carried out in\cite{georgi}. The
gluon-gluon couple to the higgs boson through the quark loop is
shown in Fig.\ref{fig:gg2h}. Although the higher order corrections
by
QCD\cite{Graudenz,hproduction,Marzani:2008az,NNLO,NNLO1,NNLO2,NNNLO}
and electroweak\cite{EWC} to the process have been calculated and
discussed, here we only take the lowest order for our present
purpose as the parameters in 2HDM have not well been constrained and
the uncertainty remains large. To lowest order, the parton cross
section can be expressed as\cite{hproduction,Marzani:2008az}:
\begin{eqnarray}
\hat{\sigma}(gg\to H)&=&\frac{\alpha_s^2 G_f}{128\sqrt{2}\pi}
\left|\sum_f \frac{y_f v}{m_f} A_f(\tau_f)\right|^2\nonumber\\
&=&\frac{\alpha_s^2}{256\pi}\left|\sum_f \frac{y_f}{m_f}
A_f(\tau_f)\right|^2,
\end{eqnarray}
where $y_f $ is the Yukawa coupling of $f$-quark. The scaling
variable is defined as
\begin{eqnarray}
  \tau_f=\frac{m_H^2}{4 m_f^2},
\end{eqnarray}
and the loop amplitude $A_f$ has the form
\begin{eqnarray}
A_f(\tau)=[\tau+(\tau-1)F(\tau)]/\tau^2,
\end{eqnarray}
with
\begin{eqnarray}
F(\tau)=\left\{
\begin{array}{cc}
  \arcsin^2\sqrt{\tau}, & \tau\le 1, \\
  -\frac{1}{4}\left[\log\frac{\sqrt{\tau}+\sqrt{\tau-1}}{\sqrt{\tau}-\sqrt{\tau-1}} -i\pi\right]^2, &
  \tau>1.
\end{array}\right. \label{eq:FT}
\end{eqnarray}
Namely if the Higgs mass is smaller than the threshold of the
$f$-quark pair production, the amplitude is real, while above the
threshold, the amplitude becomes complex according to the Cutkosky
rule. As the Yukawa coupling of top-quark is much larger than the
Yukawa coupling of bottom quark in SM, the Higgs production is
mainly through the triangular loop of top quark, and the
contributions from other quarks can be omitted. While for the new
neutral Higgs boson in 2HDM, the situation can be different due to
the possible large Yukawa couplings of the $b$-quarks such as shown
in Eq.(\ref{eq:case}).

The running strong coupling $\alpha_s(\mu)$ is known to be as
follows\cite{pdg}:
\begin{eqnarray}
\alpha_s(\mu)=\frac{\alpha_s(m_t)}{v(\mu)}\left(1-\frac{\beta_1}{\beta_0}
\frac{\alpha_s(m_t)}{4\pi v(\mu)} \ln v(\mu)\right),
\end{eqnarray}
where:
\begin{eqnarray}
  \beta_0&=&\frac{11 N-2f}{3},\\
  \beta_1&=&\frac{34}{3}N^2-\frac{10}{3}N f-2 C_F f,\\
  v(\mu)&=&1-\beta_0 \frac{\alpha_s(m_t)}{2\pi} \ln\frac{m_t}{\mu},
\end{eqnarray}
with $\alpha_s(m_t=174\GeV)=0.108$. This formula is valid at
$\mu>m_b$ with $f=5$ when $m_b<\mu<m_t$ and $f=6$ when $\mu>m_t$.

The corresponding hadronic cross section $\sigma$ can be obtained by
convolution with the gluon-gluon luminosity $\mathcal{L}(\omega)$:
\begin{eqnarray}
\sigma(PP\to H)&=&\int dx_1 dx_2 g_1(x_1)g_2(x_2) \hat{\sigma}_{gg}(\hat{s}=x_1\cdot x_2 S)\nonumber\\
&=&\int d \omega \hat{\sigma}_{gg}(\hat{s}=\omega\cdot S) \int\frac{dx_2}{x_2} g_1(\frac{\omega}{x_2}) g_2(x_2)\nonumber\\
&\equiv&\int d \omega \hat{\sigma}_{gg}(\hat{s}=\omega\cdot S)\mathcal{L}(\omega),
\end{eqnarray}
where $S=2P_1\cdot P_2$ is the center-of-mass energy of the
proton-proton collisions, and $g_i$ is the Parton Distribution
Function (PDF, \cite{cteq}) of the gluon from the $i$-th incoming
proton.

\begin{figure}[htb]
  \includegraphics[scale=0.5]{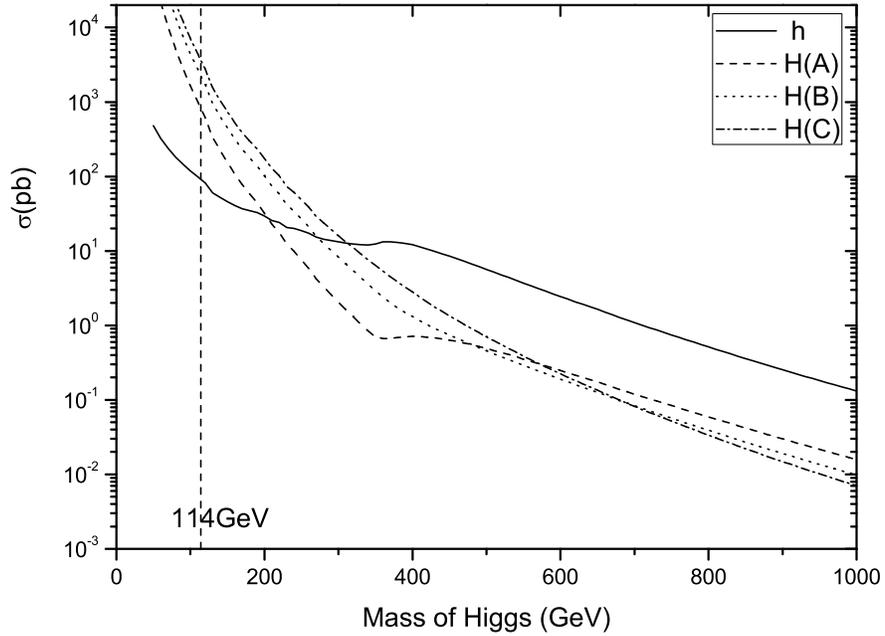}
\caption{The production cross section of the neutral Higgs. The
solid line is the result of the SM-like Higgs $h$, and the A, B, C
three lines are the results of $H$ with different Yukawa couplings.}
   \label{fig:pph}
\end{figure}

We have shown in Fig.\ref{fig:pph} the production cross-section on
LHC for the neutral Higgs bosons with the three typical Yukawa
couplings mentioned in the previous section (Eq.(\ref{eq:case})). Where $h$
denotes the SM-like Higgs in the 2HDM and H the new Higgs boson in
the 2HDM. From the figure one can see that when the Higgs masses are
light ($< 200$ GeV), the production cross-section for $H$ is larger
than the one for $h$. The reason is that at the lower mass the light
quark contributions to the Higgs $H$ production can not be omitted
for the possible large Yukawa couplings. When the Higgs mass goes to
be heavy, the loop contributions from top quark become dominant, as
the top quark Yukawa coupling of $H$ is smaller than the one of $h$,
thus the production cross section of $H$ is smaller than the one of
$h$ when they are heavy.

Note that as the SM Higgs is similar to $h$ in the 2HDM when
neglecting the possible mixing among neutral Higgs bosons, therefore
it is hard to distinguish with the SM Higgs, unless the charged
Higgs is very light, so that $h$ in the 2HDM can decay to $H^+ H^-$,
while the vertex comes from the Higgs potential which is strongly
model-dependent. In this note, we will not consider the possible
charged Higgs effect to the neutral Higgs decay modes. We are going
to pay attention to the neutral Higgs mixing effect which also make
the $h$ in the 2HDM differ to the SM Higgs. It is seen that when the
masses of $h$ and $H$ are both at $200\GeV\sim300\GeV$, the cross
section of $h$ and $H$ are similar, but it will be shown below that
they have very different decay modes.

In general, the Yukawa couplings of the new neutral Higgs bosons can
be complex, namely one can write $\xi^{q}=|\xi^{q}| e^{i\delta_q}$,
so that the Yukawa interactions of the neutral Higgs bosons in
Eq.(\ref{Yukawa}) can be written into two parts:
\begin{eqnarray}
\mathcal{L}_{Y}&=&\bar{U}\frac{\xi^U}{\sqrt{2}}\frac{1-\gamma^5}{2}
U\phi_2^0+\bar{U}\frac{{\xi^U}^\dag}{{\sqrt{2}}}
\frac{1+\gamma^5}{2}U\phi_2^0+(U\to D)+...\nonumber\\
                &=&\frac{|\xi^q|}{\sqrt{2}}\cos\delta_q\bar{q}q\phi_2-i\frac{|\xi^q|}{\sqrt{2}}\sin\delta_q\bar{q}\gamma^5q\phi_2+...,
\end{eqnarray}
where the second part leads to additional contributions through a
kind of anomaly loop diagrams,  the resulting amplitude has the
following form\cite{gamma5,wugamma5}:
\begin{eqnarray}
\mathcal{M}^{A}=\sum_f\frac{ig_s^2|\xi^{f}|\sin\delta_f\epsilon_{\mu\nu\rho\sigma}\epsilon^{a\mu}_1
\epsilon_2^{b\nu}k_1^\rho k_2^\sigma \delta ^{ab}}{4{\sqrt{2}}\pi^2
M_H}B_f(\tau_f),
\end{eqnarray}
where the $a$ and $b$ are the color index and the function
$B_f(\tau_f)$ is given by
\begin{eqnarray}
  B_f(\tau_f)=\frac{F(\tau_f)}{\sqrt{\tau_f}},
\end{eqnarray}
where $F(\tau)$ is defined as Es.(\ref{eq:FT}).

\begin{figure}
  \includegraphics[scale=0.25]{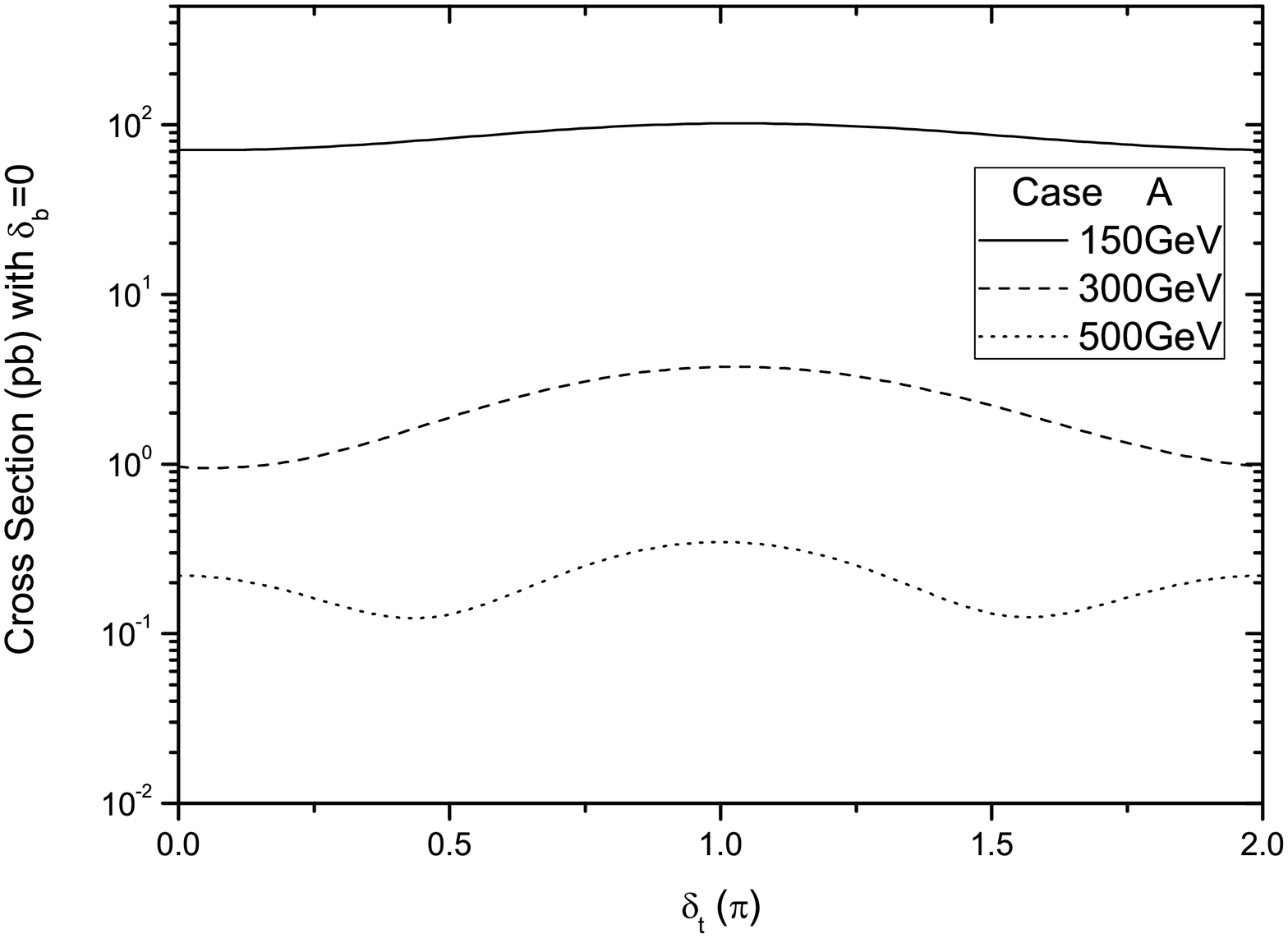}
  \includegraphics[scale=0.25]{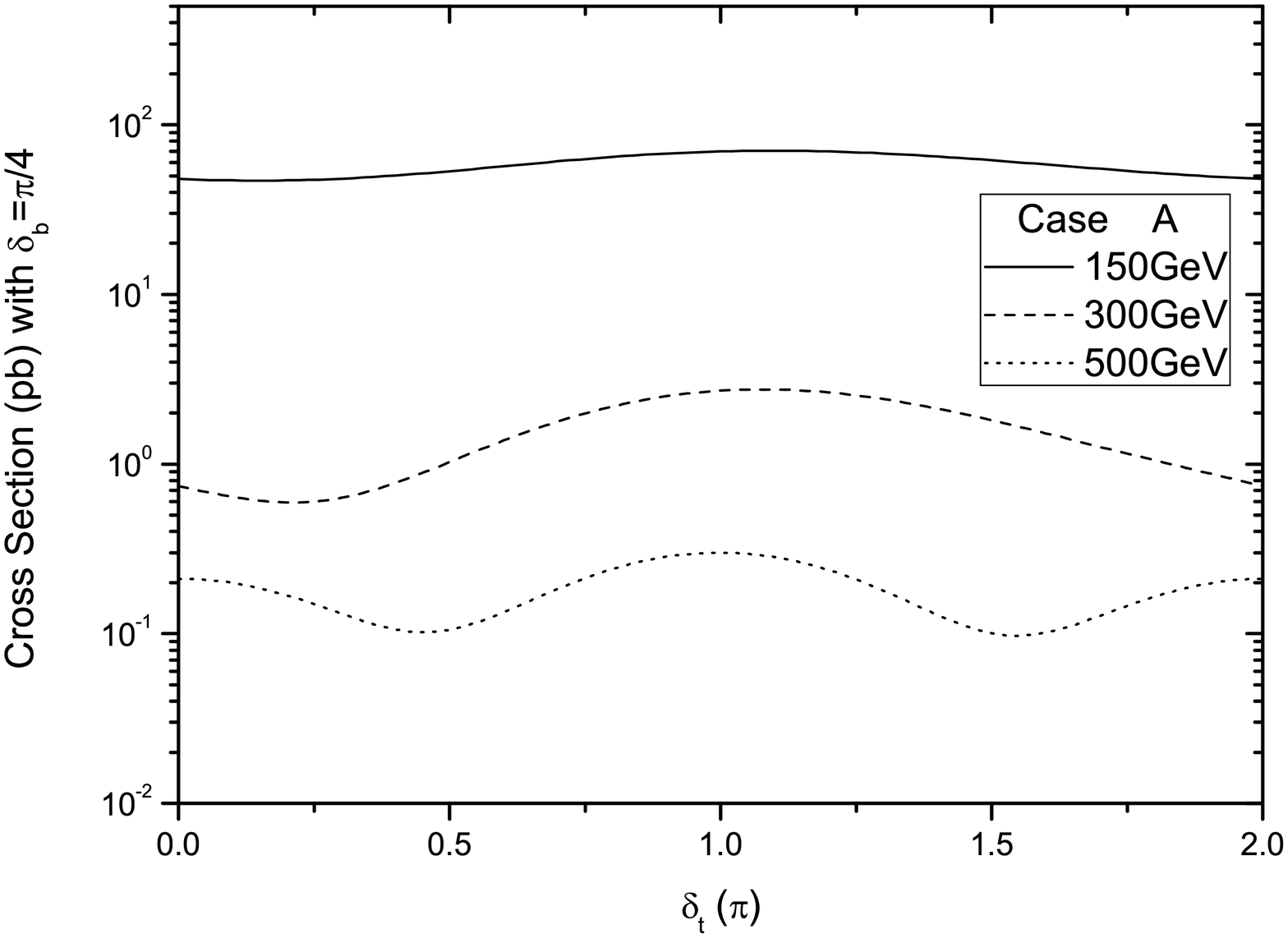}\\
  \includegraphics[scale=0.25]{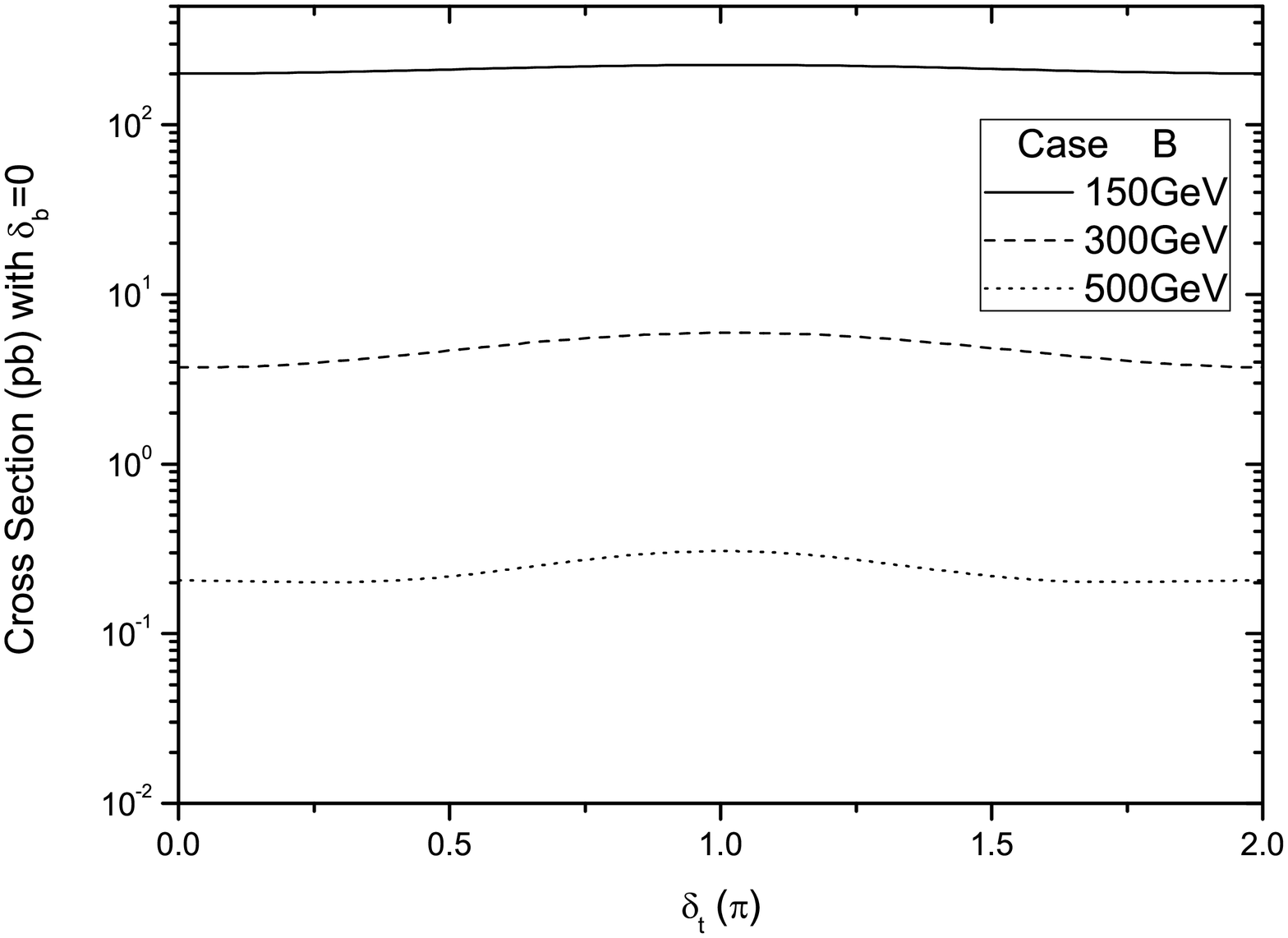}
  \includegraphics[scale=0.25]{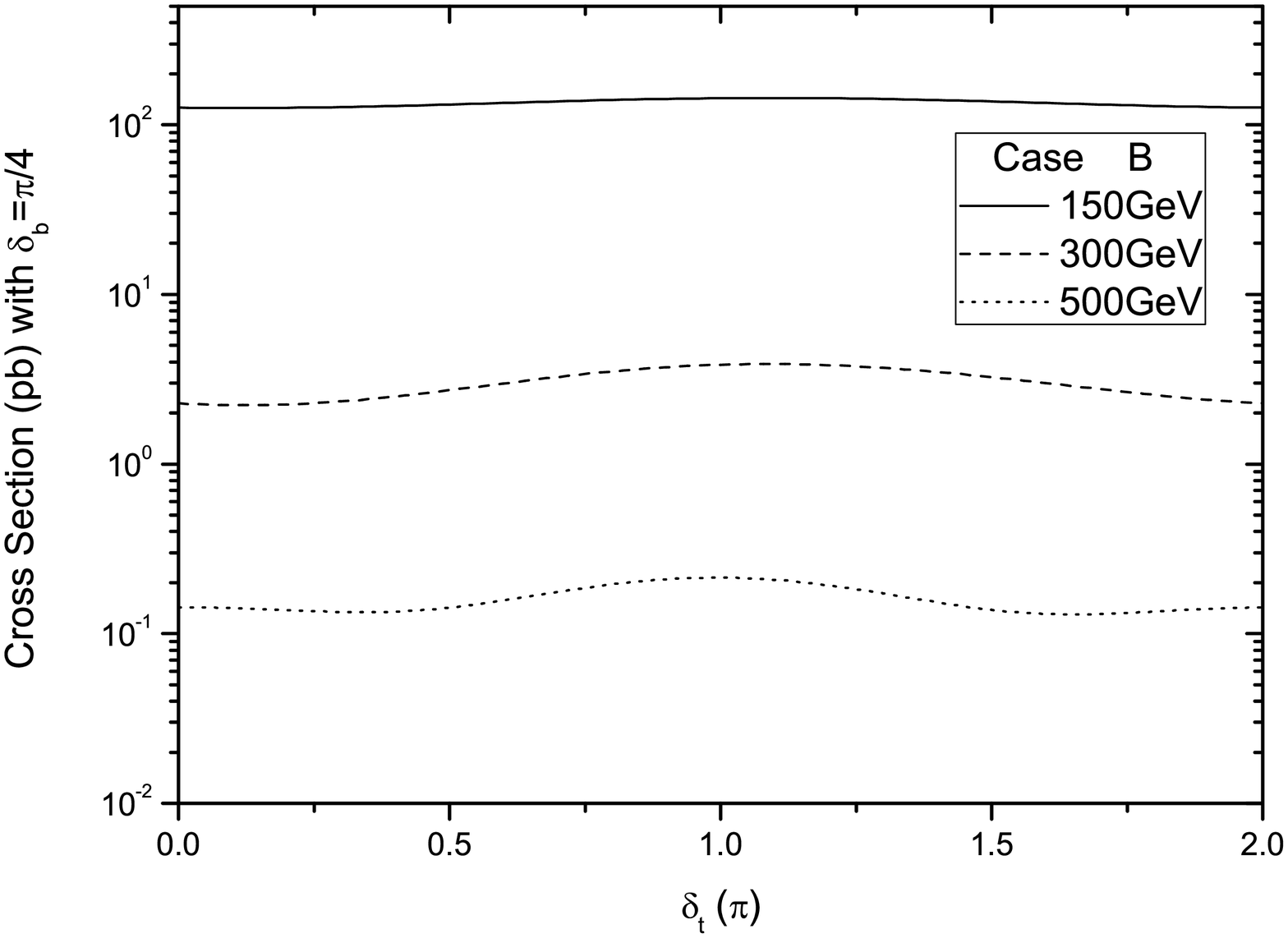}\\
  \includegraphics[scale=0.25]{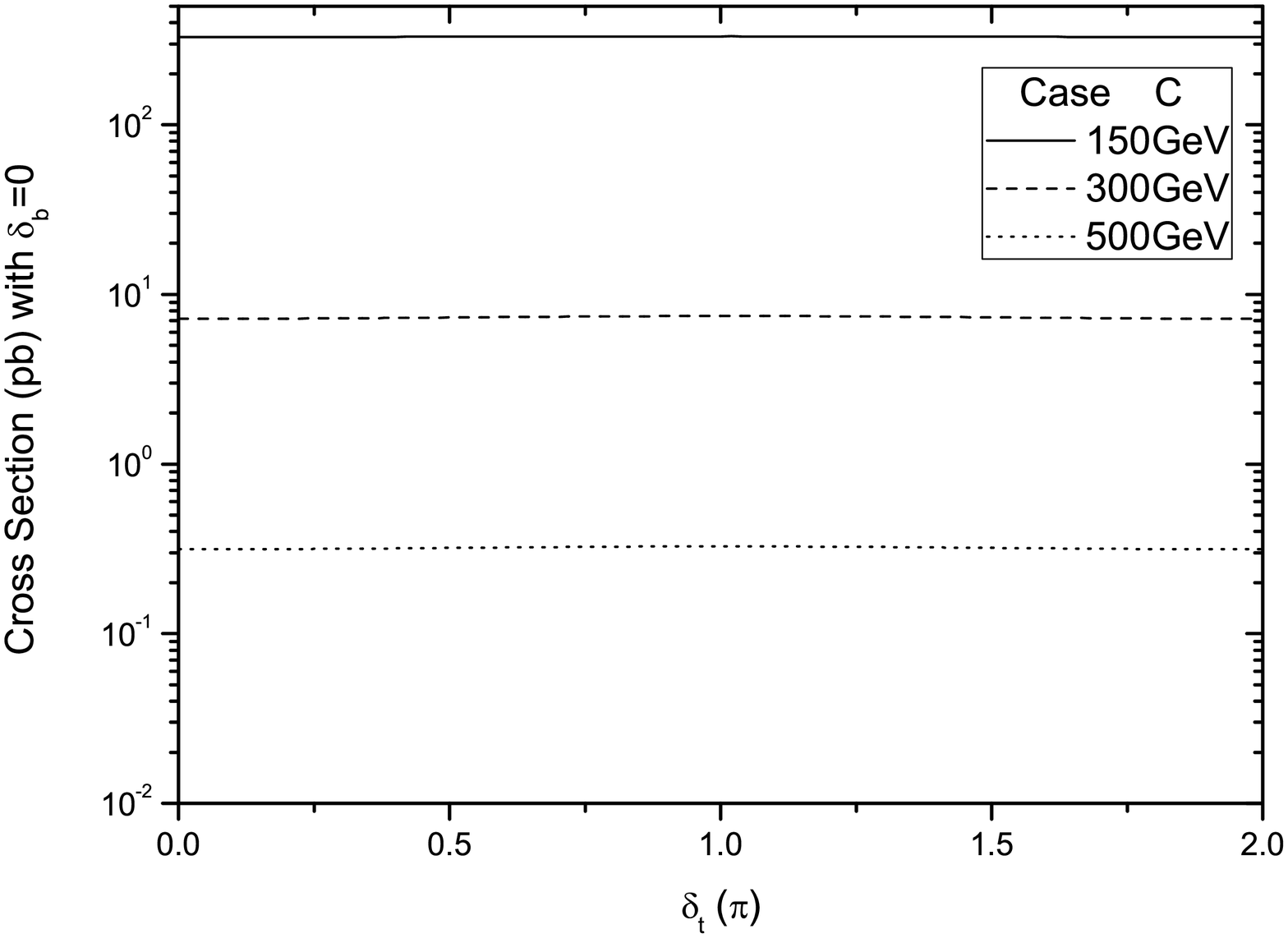}
  \includegraphics[scale=0.25]{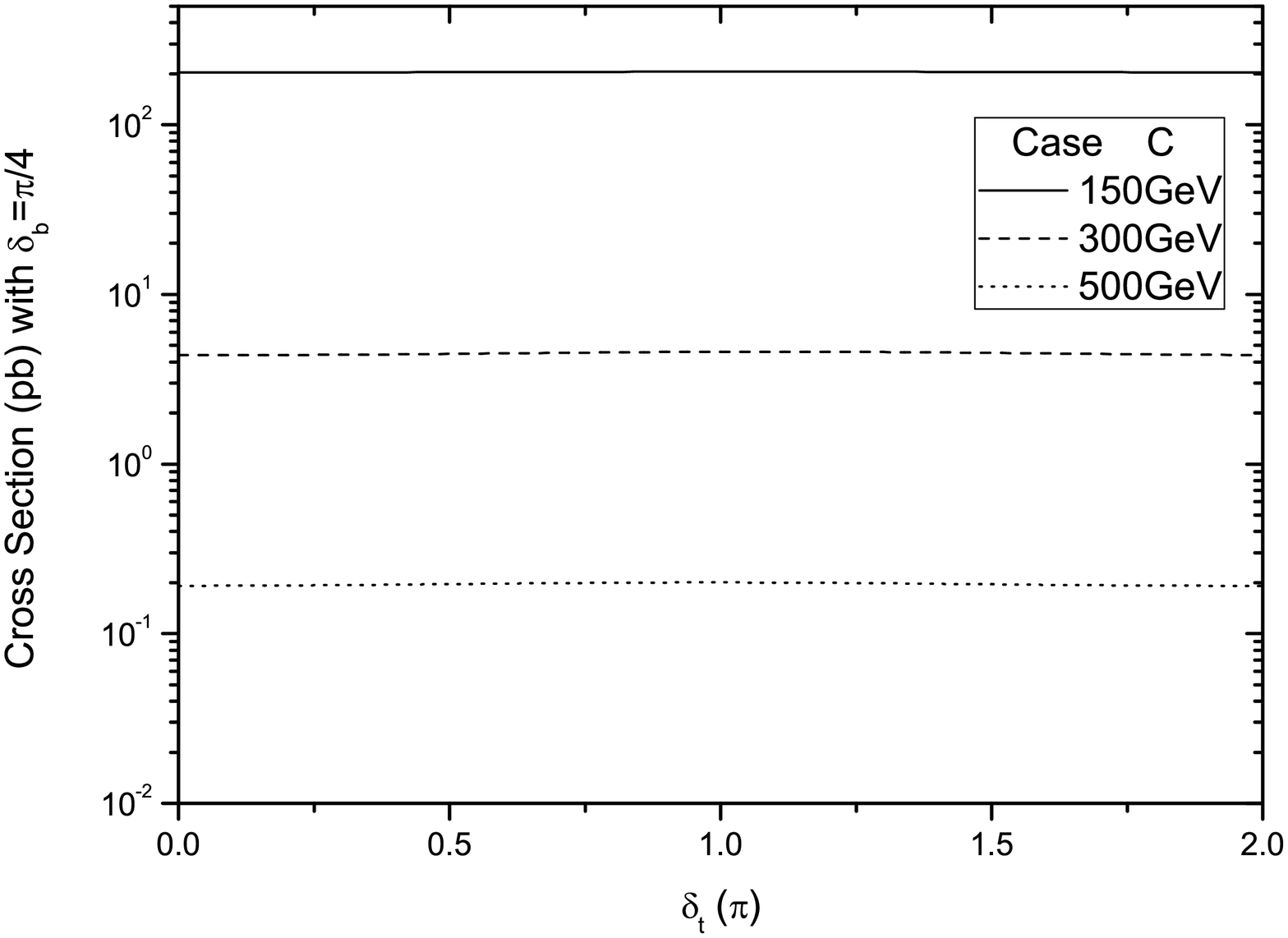}
\caption{The production cross section depends on the phase of the
top-quark Yukawa coupling phase $\delta_t$ with the three typical
absolute values given in Eq (\ref{eq:case}) and different Higgs
masses. The results for $\delta_b=0$ are listed on the left side,
and $\delta_b=\pi/4$ on the right side.}
  \label{fig:complex_top}
\end{figure}

\begin{figure}[htb]
    \includegraphics[scale=0.25]{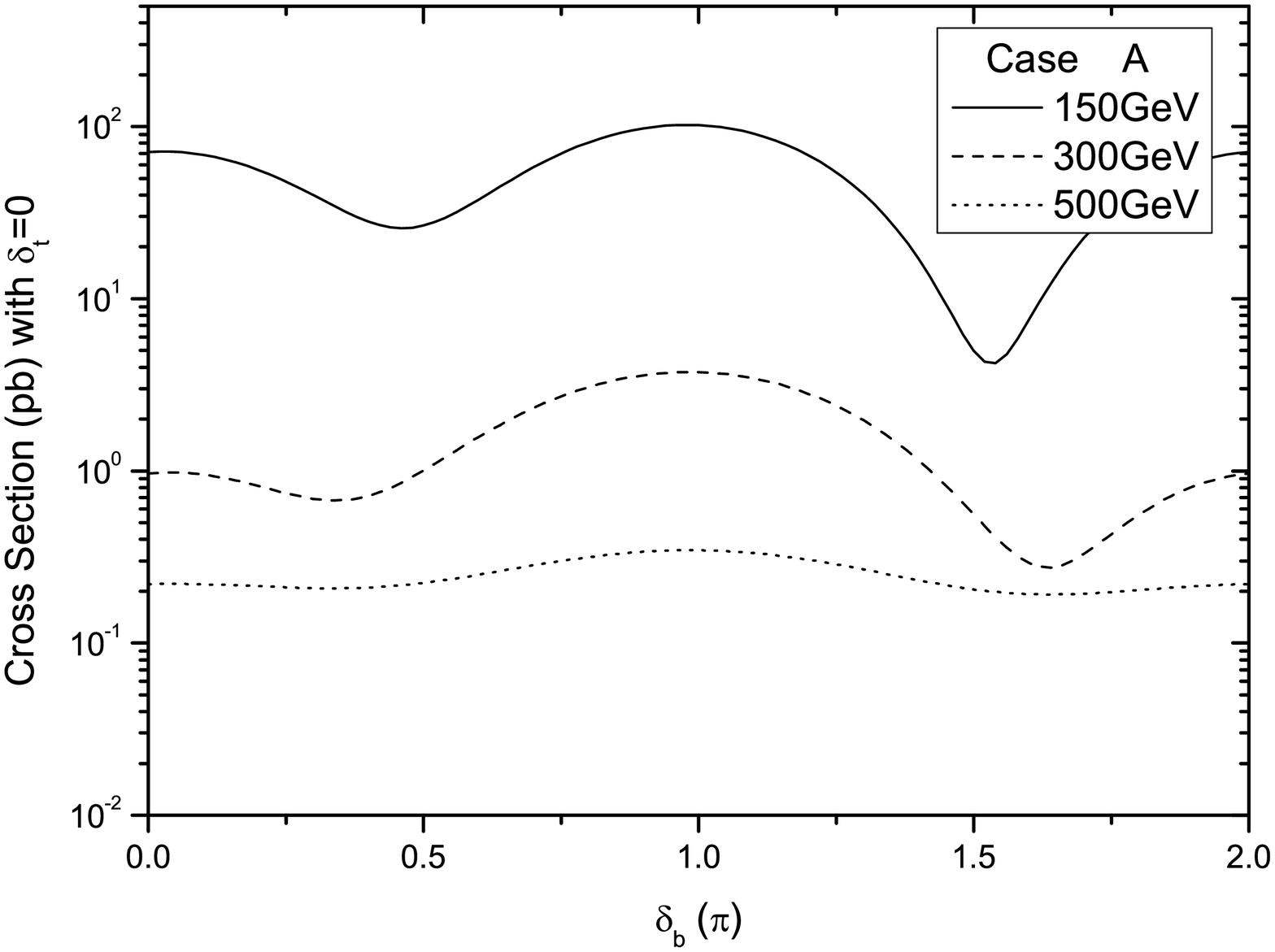}
  \includegraphics[scale=0.25]{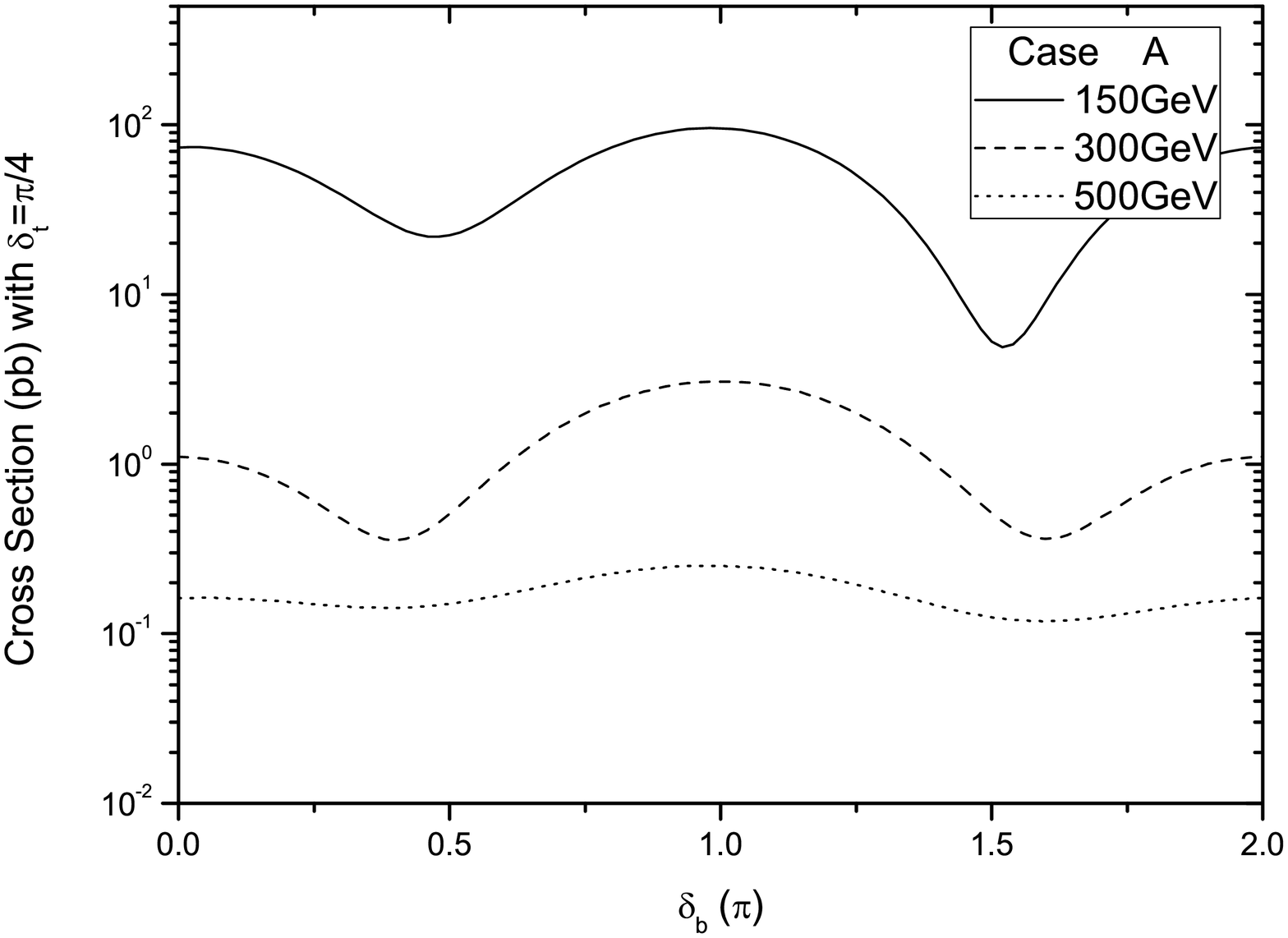}\\
  \includegraphics[scale=0.25]{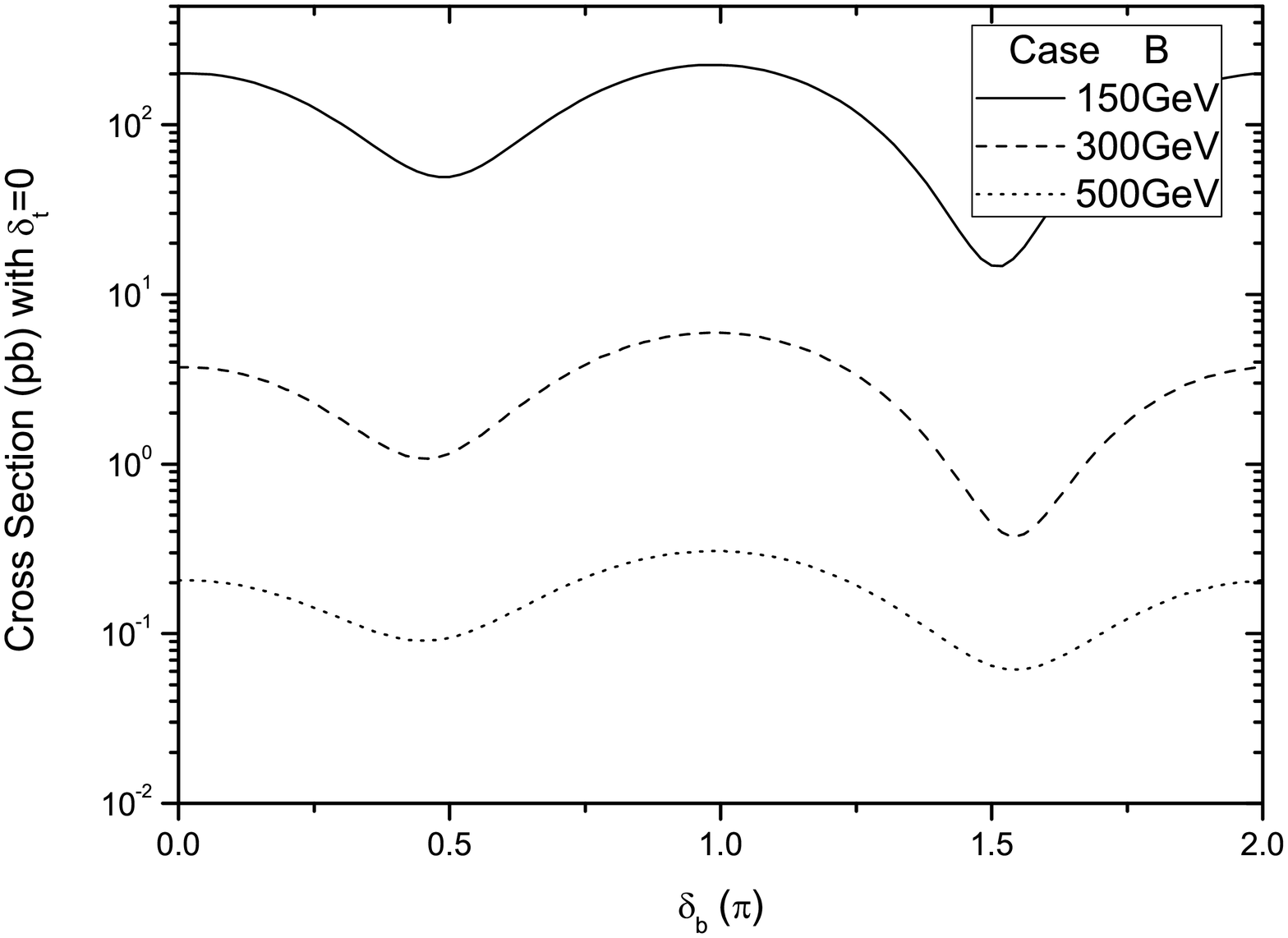}
  \includegraphics[scale=0.25]{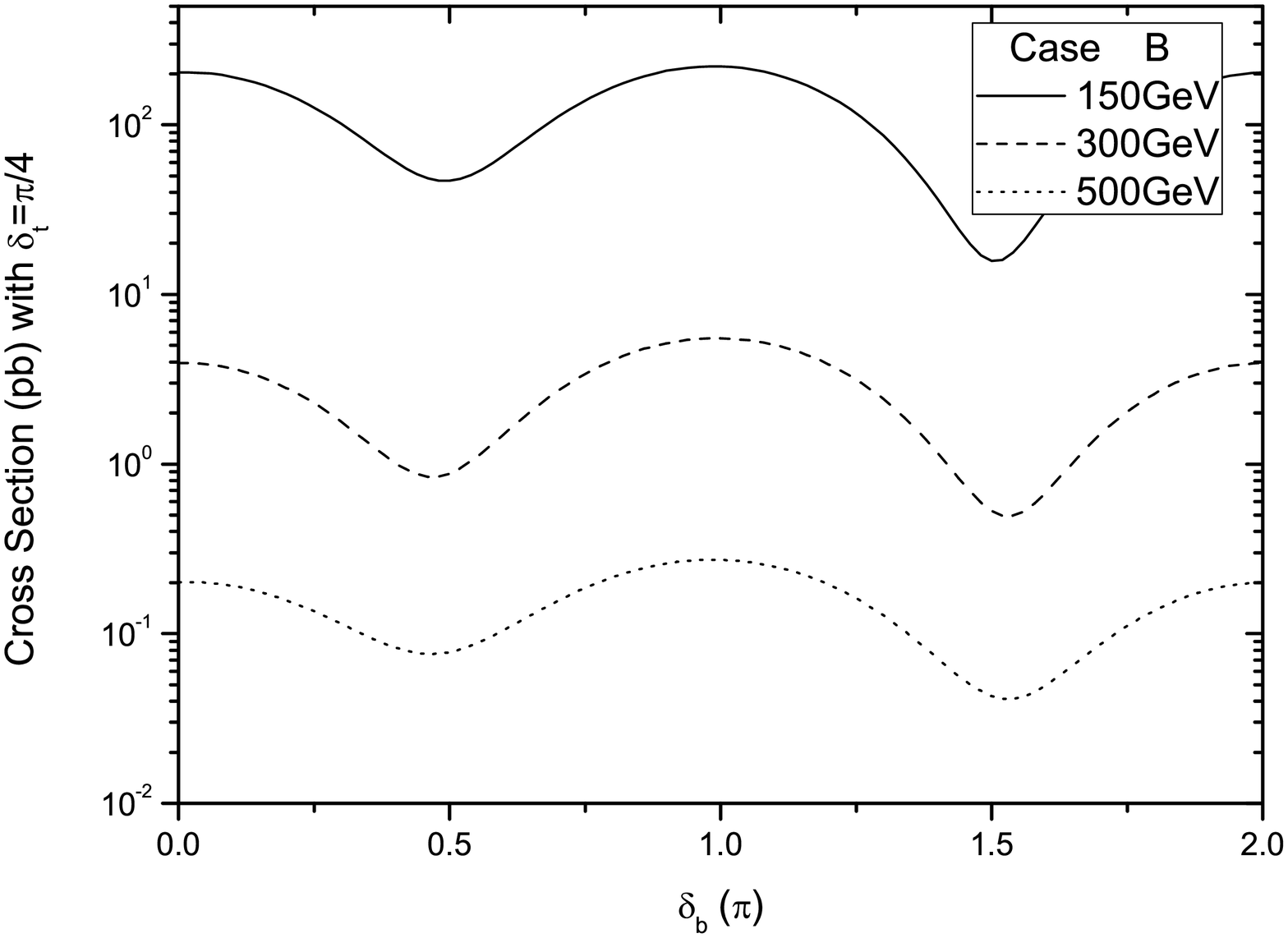}\\
  \includegraphics[scale=0.25]{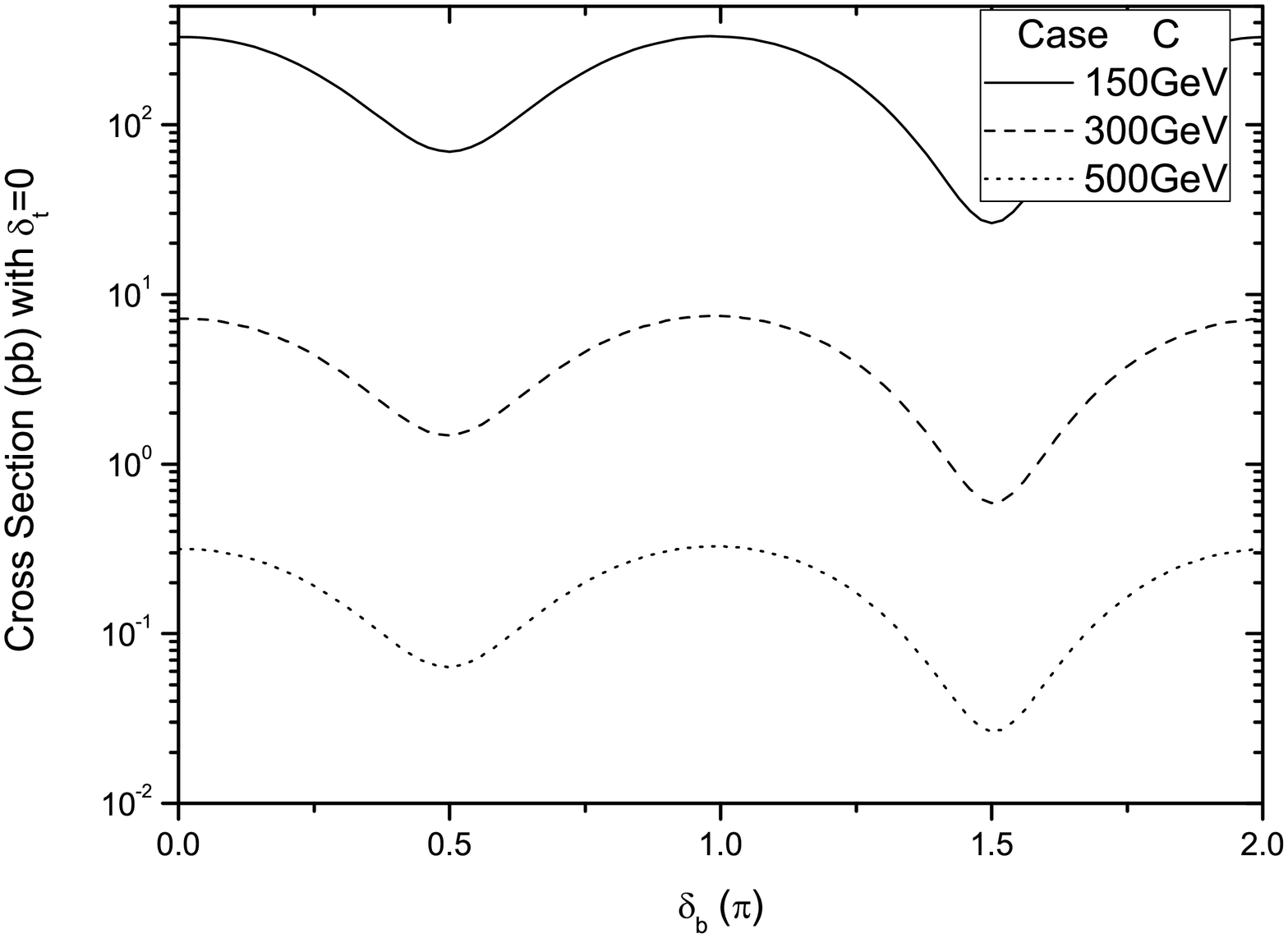}
  \includegraphics[scale=0.25]{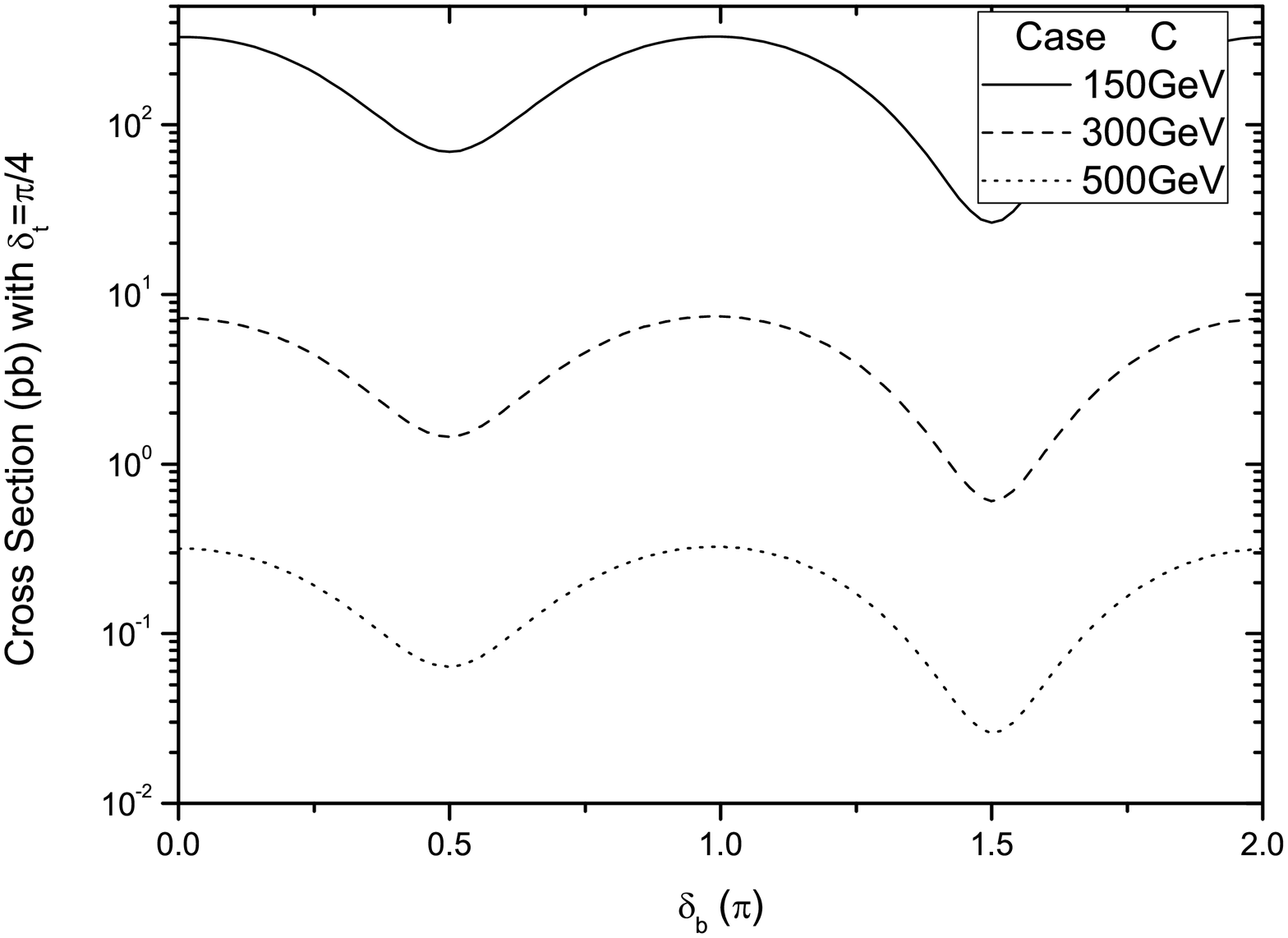}
\caption{The production cross section depends on the phase of the
bottom-quark Yukawa coupling phase $\delta_b$ with the three typical
absolute values given in Eq (\ref{eq:case}) and different Higgs
masses. The results for $\delta_t=0$ are listed on the left side,
and $\delta_t=\pi/4$ on the right side.} \label{fig:complex_bottom}
\end{figure}

The cross section is calculated as the function of $\delta_t$ with
$\delta_b=0$ and $\delta_b =\pi/4$ plotted in
Fig.\ref{fig:complex_top}, and as the function of $\delta_b$ with
$\delta_t=0$ and $\delta_t =\pi/4$ plotted in
Fig.\ref{fig:complex_bottom}, where we have also considered three
typical cases given in Eq.(\ref{eq:case}) with the absolute values
of the Yukawa couplings. It can be seen from the figures that: For
the case A, the effect of the phase in top quark Yukawa coupling
becomes significant when the mass of the Higgs $H$ increases, which
is very different from the case with real Yukawa couplings and
implies that the top quark loop contribution to the cross-section in
gluon-gluon fusion is dominant for a heavy Higgs $H$, while for a
light Higgs $H$, the bottom quark contribution becomes important.
However, the situation becomes different for the case B and case C
when the top quark Yukawa coupling goes to be small, especially for
the case C where the bottom quark loop contribution becomes
dominant, and the phase of the top quark Yukawa coupling has less
effect to the cross-section.

\section{The Higgs decays\label{section:decays}}

\begin{figure}
  \includegraphics[scale=0.5]{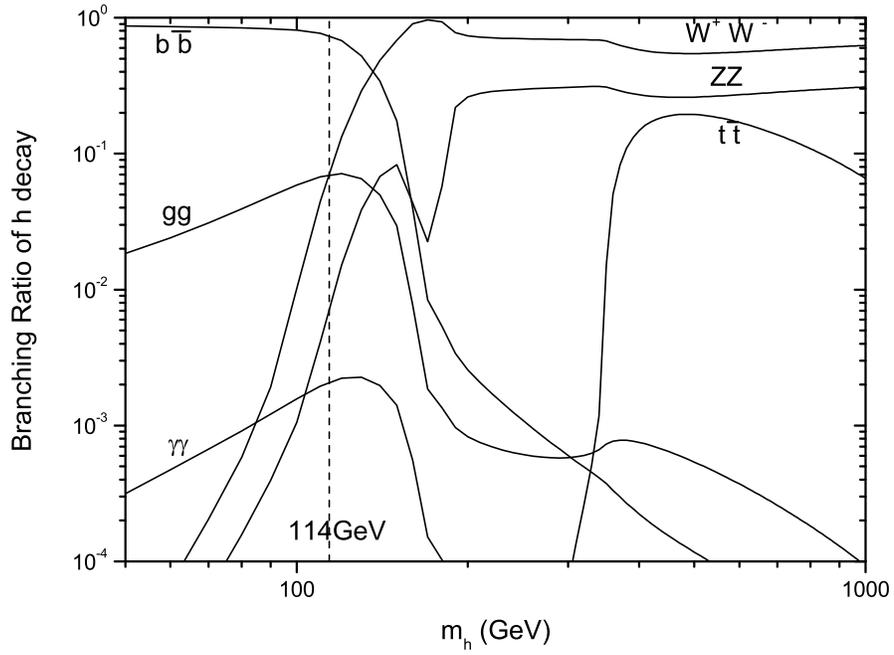}
\caption{The decay of the Higgs $h$ in 2HDM without mixing, which
likes the SM Higgs.}
  \label{fig:smdecay}
\end{figure}

As the neutral Higgs bosons cannot be directly detected, they are
probed only through the final states of their decays. For the
SM-like Higgs $h$ in the general 2HDM, without considering the
mixings among the neutral Higgs, it has the same Yukawa coupling as
the Higgs in SM at tree level (except the couplings with other
Higgs). As shown in Fig.\ref{fig:smdecay}, which likes the SM Higgs
decay\cite{hdecay}, when its mass is lower than the WW threshold,
the $b\bar{b}$ pair production is a dominant decay mode. However,
the process $h\to\gamma\gamma$ is also sizable as shown in
Fig.\ref{fig:smdecay}, which is known to be a golden channel for
detecting the light neutral Higgs due to the clean background.

It can also be seen from Fig.\ref{fig:smdecay} that if $M_h > 160$
GeV there are two dominant processes concerning the W and Z bosons,
where the W and Z bosons can decay to quarks and leptons. They are
the golden channels for searching the heavy neutral Higgs $h$. The
situation is very different for the new neutral Higgs $H$. After the
transformation of the Higgs in Eq.(\ref{eq:transform}), the gauge
part of the Higgs in the new basis can be written as \cite{YLW1}
\begin{eqnarray}
  \mathcal{L}_{\rm G}&=&\left(D_\mu H_1\right)^\dag\left(D^\mu
  H_1\right)+\left(D_\mu H_2\right)^\dag\left(D^\mu
  H_2\right)\nonumber\\
  &=&\frac{1}{2}\partial_\mu\phi^0_1\partial^\mu\phi^0_1+\frac{(v+\phi^0_1)^2}{8}\left[(g\prime^2+g^2)Z^2+2g^2W^+W^-\right]\nonumber\\
  &&+\frac{1}{2}\left(\partial\phi_2^0\partial^\mu\phi^0_2+\partial\phi_3^0\partial^\mu\phi^0_3\right)+\partial_\mu
  H^-\partial^\mu H^+\nonumber\\
  &&+e^2
  H^+H^-A^2+\frac{(g^2-g^{\prime2})^2}{4(g^2+g^{\prime2})}H^+H^-Z^2+\frac{g^2}{2}H^+H^-W^+W^-+\frac{e(g^2-g^{\prime2})}{\sqrt{g^2+g^{\prime2}}}H^+H^-Z\cdot A\nonumber\\
  &&+\frac{g^2}{4}W^+W^-(\phi_2^{02}+\phi_3^{02})+\frac{g^2+g^{\prime2}}{8}(\phi_2^{02}+\phi_3^{02})Z^2\nonumber\\
  &&+\left\{ieA^\mu H^-\partial_\mu H^++i\frac{g^2-g^{\prime2}}{2\sqrt{g^2+g^{\prime2}}}Z^\mu H^-\partial_\mu H^++\frac{ig}{2}W^-_\mu(\phi_2^0-i\phi_3^0)\partial^\mu H^+\right.\nonumber\\
  &&+\frac{eg}{2}A^\mu W^-_\mu H^+(\phi^0_2-i\phi_3^0)+\frac{g}{4}\frac{g^2-g^{\prime2}}{\sqrt{g^2+g^{\prime2}}}H^+W^-_\mu Z^\mu(\phi_2^0-i\phi_3^0)\nonumber\\
  &&+\frac{ig}{2}H^-W^+_\mu(\partial^\mu\phi_2^0+i\partial^\mu\phi_3^0)+\frac{i\sqrt{g^2+g^{\prime2}}}{4}(\phi^0_2-i\phi^0_3)Z^\mu(\partial_\mu\phi^0_2+i\partial_\mu\phi^0_3)\nonumber\\
  &&\left.-\frac{g\sqrt{g^2+g^{\prime2}}}{4}H^-W^+_\mu
  Z^\mu(\phi_2^0+i\phi_3^0)+H.c.\right\}.
\end{eqnarray}
It is seen that without considering the mixing among the neutral
Higgs bosons, namely the neutral Higgs gauge interaction eigenstates
$(\phi_1^0,\phi_2^0,\phi_3^0)$ are the same as the mass eighenstates
$(h,H,A)$, there are no direct $WWH$ and $ZZH$ interactions, this is
because the vacuum expectation value of the Higgs doublet $H_2$ in a
rotating basis vanishes, $\langle H_2\rangle=0$. In this special
case, the Higgs $H$ cannot decay to $WW$ and $ZZ$ at the tree level,
thus the $f\bar{f}$ channels are always the dominant decay modes of
$H$ for its whole mass range. As there is no symmetry to forbid the
mixing among the neutral Higgs bosons, in general we shall consider
the mixing among the Higgs bosons which will be discussed later on.

\subsection{The $\gamma\gamma$, $WW$ and $ZZ$ Modes}

\begin{figure}
  \includegraphics[scale=0.5]{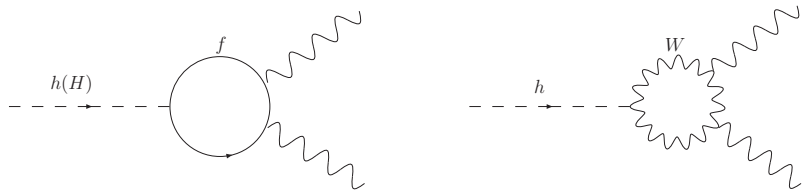}
\caption{The Feynman diagrams for the decay of the Higgs to
$\gamma\gamma$. For $H\to\gamma\gamma$ only the fermion-loop
contributes. }
  \label{fig:hgmgm}
\end{figure}

\begin{figure}[htb]
  \includegraphics[scale=0.5]{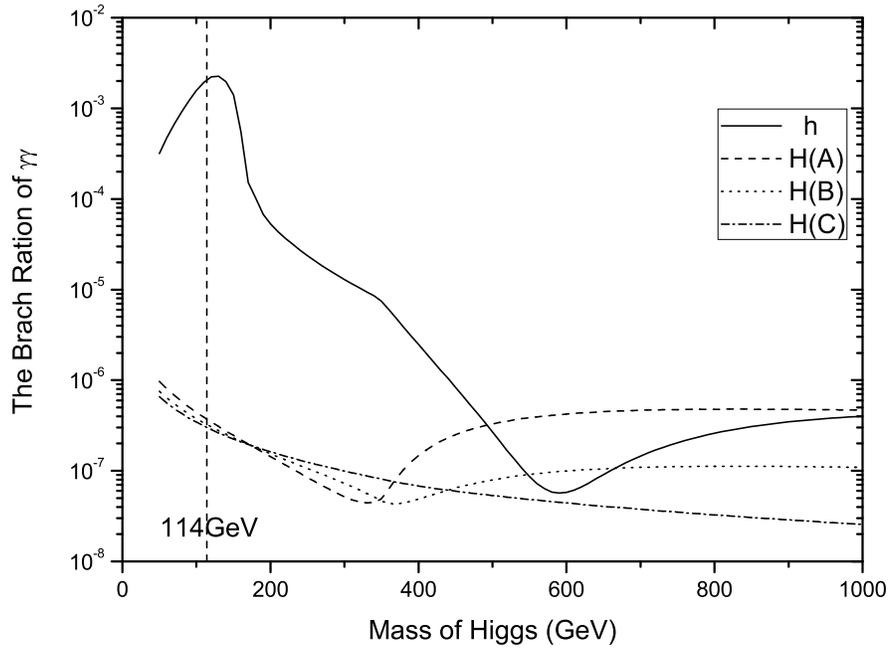}
\caption{The Higgs decay to $\gamma\gamma$ as the function of Higgs
mass. The solid line is for the SM-like Higgs $h$, other three lines
for the new Higgs $H$ with three different Yukawa couplings given in
Eq.(\ref{eq:case}).}
  \label{fig:h2gmgm}
\end{figure}

The SM-like neutral Higgs $h$ with mass M decays to $\gamma\gamma$
through the fermion-loop and W-loop as shown in Fig.\ref{fig:hgmgm},
the decay width is given by:
\begin{equation}
\Gamma(\gamma\gamma)=\frac{G_F\alpha^2
M^3}{128\sqrt{2}\pi^3}\left|\sum_f N_c Q_f^2\frac{y_f v}{m_f}
A_f(\tau_f)+A_W(\tau_W)
  \right|^2,
\label{eq:gmgm}
\end{equation}
where the color factor $N_c$ is 3 for quarks and 1 for
leptons. The amplitude $A_W$ is contributed from the W-loop
\cite{hproduction}:
\begin{eqnarray}
A_W(\tau)=-\left[2\tau^2+3\tau+3(2\tau-1) F(\tau)\right]/\tau^2.
\end{eqnarray}
As discussed in \cite{hproduction}, the W-loop contribution is
dominated when the Higgs mass is below 600 GeV. For the new neutral
Higgs decay $H\to \gamma\gamma$, its decay width is smaller than the
one of $h$ when the Higgs mass is below 600 GeV , this is because
there is no W-loop contribution to  $H\to \gamma\gamma$. While the
total decay width of $H$ can be larger than the one of $h$ as the
$H\to b\bar{b}$ can be dominated and much larger than $h\to
b\bar{b}$ due to a possible larger Yukawa coupling. Thus the
Branching Ratio of $H\to \gamma\gamma$ is very small in comparison
with Br($h\to \gamma\gamma$) as shown in Fig.\ref{fig:h2gmgm}.

If the Higgs boson is very heavy, the $WW$, $ZZ$, $t\bar{t}$
channels open and become dominant. The fractional widths of the
SM-like Higgs $h$ decaying to $WW$ and $ZZ$ at tree level are given
by\cite{pdg}:
\begin{eqnarray}
  &&\Gamma(WW)=\frac{G_f M^3 \beta_W}{32\pi\sqrt{2}}(4-4a_W+3a_W^2),\\
  &&\Gamma(ZZ)=\frac{G_f M^3 \beta_Z}{64\pi\sqrt{2}}(4-4a_Z+3a_Z^2),
\end{eqnarray}
where $a_W=1-\beta_W^2=4 m_W^2/M^2$ and $a_Z=1-\beta_Z^2=4
m_Z^2/M^2$ with $M$ is mass of the SM-like Higgs.

Note that the new neutral Higgs $H$ cannot decay to $WW$ and $ZZ$ at
tree level without considering the mixing with the Higgs $h$ which is
going to be discussed in section \ref{section:mixings}.

\subsection{$H/h \to \bar{f}f$ Decay}

\begin{figure}[htb]
  \includegraphics[scale=0.5]{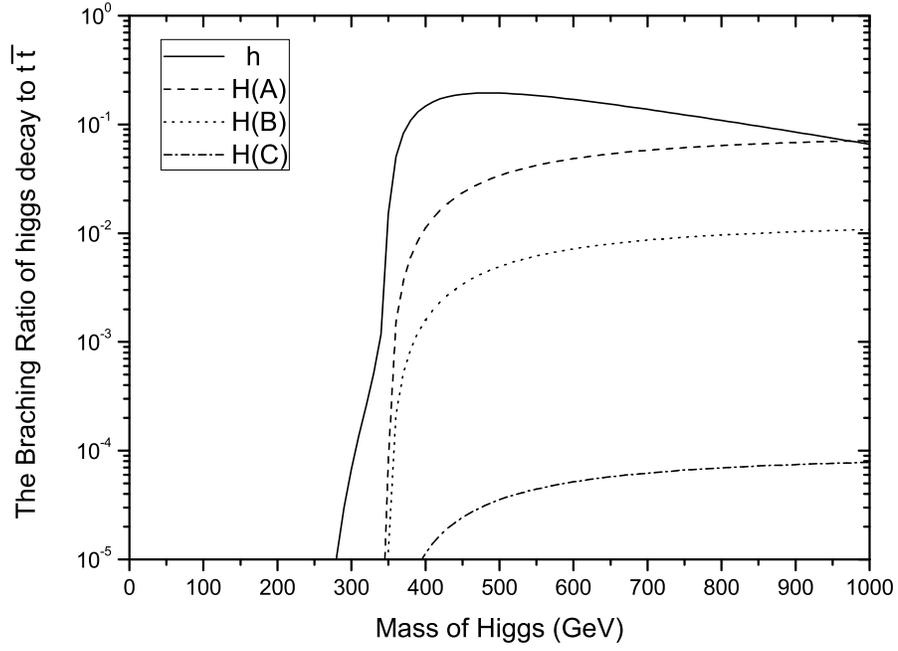}
  \includegraphics[scale=0.5]{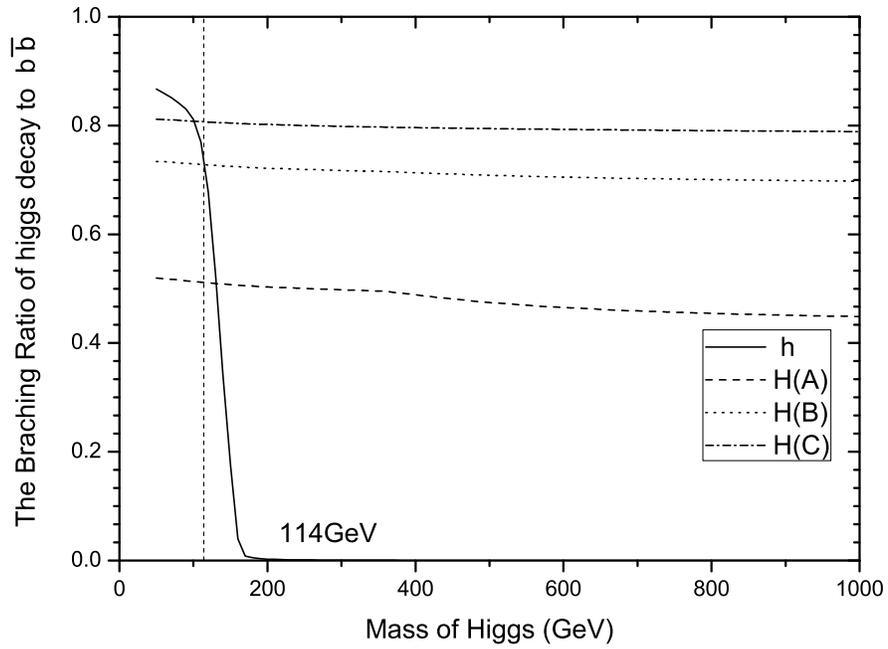}
\caption{The Higgs decay to $t\bar{t}$ and $b\bar{b}$ as the
function of the Higgs mass. The solid line is for the SM-like Higgs
$h$, other three lines for the new Higgs $H$ with three different
Yukawa couplings given in Eq.(\ref{eq:case}).}
  \label{fig:h2ff}
\end{figure}

The fermion decay modes are dominated for both $h$ and $H$ when
their masses are below the threshold of $WW$ pair production. At
tree level the fractional width of the higgs (both $h$ and $H$)
decays to fermion-antifermion pair is given by
\begin{equation}
  \Gamma(\bar{f}f)=N_c\frac{y_f^2}{4\pi M^2}(M^2-4m_f^2)^{3/2},
  \label{eq:ffbar}
\end{equation}
which shows that when the Higgs mass $M$ is much larger than the
fermion mass $m_f$, the decay width is proportional to the Higgs
mass. For the new neutral Higgs $H$, all other channels can in
general be omitted in comparing with the $f\bar{f}$ channels when
neglecting the large mixing effects between $H$ and $h$. As the $H$
can not decay to $WW$ and $ZZ$ at tree level, the branching ratio of
$f\bar{f}$ is approximately given by the ratio $N_c \xi_f^2/\sum N_c
\xi_f^2$, which is independent on the Higgs mass as shown in
Fig.\ref{fig:h2ff}. For the neutral Higgs $H$, the Yukawa coupling
of the bottom quark is taken to be larger than the one of top quark
in our present consideration, thus the $H\to b\bar{b}$ decay width
is always larger than the $H\to t\bar{t}$ decay width as shown in
Fig.\ref{fig:h2ff}. However, $H\to b\bar{b}$ is overwhelmed by the
combinatorial background from QCD b-jets production with
$\sigma(gg\to b\bar{b})\approx500\mu b$. As a matter of fact, the $H
\to b\bar{b}$ channel is now considered inaccessible at the LHC
\cite{ggtt} and it seems to be left open to study only at a next
linear collider\cite{linear,linear2}. Nevertheless, at LHC there are
some other processes to be hoped, such as the associated production
modes $W^{\pm}H(b\bar{b})$ and $Z H(b\bar{b})$\cite{associated}, and
$\gamma\gamma\to H\to b\bar{b}$\cite{gammagamma} in proton-lead (p
Pb) interactions. For the SM-like Higgs $h$, the situation is
different as the Yukawa couplings are proportional to the fermion
masses, $\xi_f\sim m_f$, its decay width to heavy quarks is larger
than the one to light quarks. And when the SM-like Higgs is lighter
than the threshold of $WW$, the $h\to b\bar{b}$ is the main decay
mode but also overwhelmed by the QCD background, so that the
$h\to\gamma\gamma$ studied as the golden channel to detect a light
Higgs.

\section{The Mixing Effects of The Neutral Higgs\label{section:mixings}}

In general, there is no symmetry to forbid the mixing between the
neutral Higgs bosons. Let us now consider the mixing between the
scalar neutral Higgs bosons $H$ and $h$, but without considering
their mixing with the pseudoscalar $A$ for simplicity. The Higgs
bosons $h$ and $H$ in the mass eigenstate are the linear
combinations of the Higgs bosons in the electroweak eigenstate
denoted in Eq.(\ref{eq:higgs})
\begin{eqnarray}
  &&h=\cos\theta \phi_1+\sin\theta \phi_2,\\
  &&H=-\sin\theta\phi_1+\cos\theta \phi_2.
\end{eqnarray}
In the mass eigenstate, the Yukawa terms in Eq.(\ref{eq:yukawa})
becomes
\begin{eqnarray} \mathcal{L}_Y&=&\bar{f}\frac{m_f}{v}f\phi_1+\frac{\xi^f}{\sqrt{2}}\bar{f}f\phi_2+...\nonumber\\
&=&m_f\left(\cos\theta\frac{1}{v}+\sin\theta\frac{\xi^f}{\sqrt{2}m_f}\right)\bar{f}fh\nonumber\\
&&+m_f\left(-\sin\theta\frac{1}{v}+\cos\theta\frac{\xi^f}{\sqrt{2}m_f}\right)\bar{f}f
H+...,
\end{eqnarray}
with $\theta\in(0,\pi)$. Note that if one renames $h$ as $H$ and $H$
as $-h$, it is the same as the replacement: $\theta+\frac{\pi}{2}$
to $\theta$ with $\theta\in(0,\frac{\pi}{2})$. Thus in the later
formulas and calculations, we only need to consider
$\theta\in(0,\frac{\pi}{2})$.

The Higgs production cross section is given by
\begin{equation}
\hat{\sigma}(gg\to
\phi)=\frac{\alpha_s^2}{256\pi}\left|\sum_f\left\{\left(\cos\theta_\phi\frac{1}{v}+\sin\theta_\phi
\frac{\xi^f}{{\sqrt{2}}m_f}\right) A_f(\tau_f)\right\}\right|^2,
\end{equation}
where $\phi$ denotes the Higgs boson $H$ or $h$. The decay widths of
Higgs boson to $\gamma\gamma$ and $ZZ$ at tree level are given by
\begin{eqnarray}
&&\Gamma(\phi\to\gamma\gamma)=\frac{G_F\alpha^2
M^3}{128\sqrt{2}\pi^3} \left|\sum_f N_c
Q_f^2\left(\cos\theta_\phi+\sin\theta_\phi\frac{\xi^f
v}{{\sqrt{2}}m_f}\right)
A_f(\tau_f)+\cos\theta_\phi A_W(\tau_W) \right|^2, \\
&&\Gamma(\phi\to ZZ)=\frac{G_f M^3
\beta_W}{64\pi\sqrt{2}}(4-4a_W+3a_W^2)\cos^2(\theta_\phi),
\end{eqnarray}
where $\theta_h=\theta$ for $\phi=h$ and
$\theta_H=\theta+\frac{\pi}{2}$ for $\phi=H$.

\begin{figure}[htb]
  \includegraphics[scale=0.35]{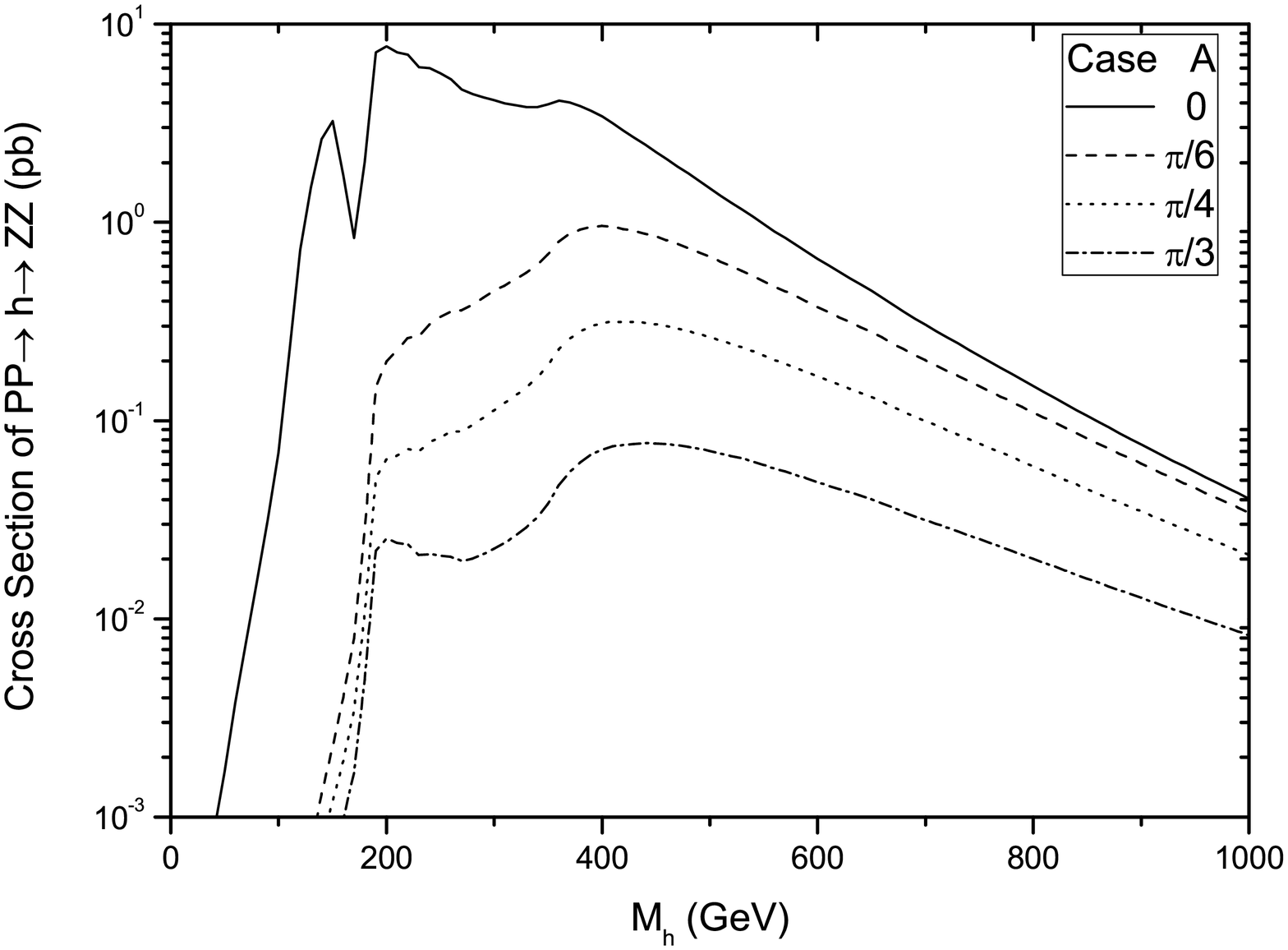}
  \includegraphics[scale=0.35]{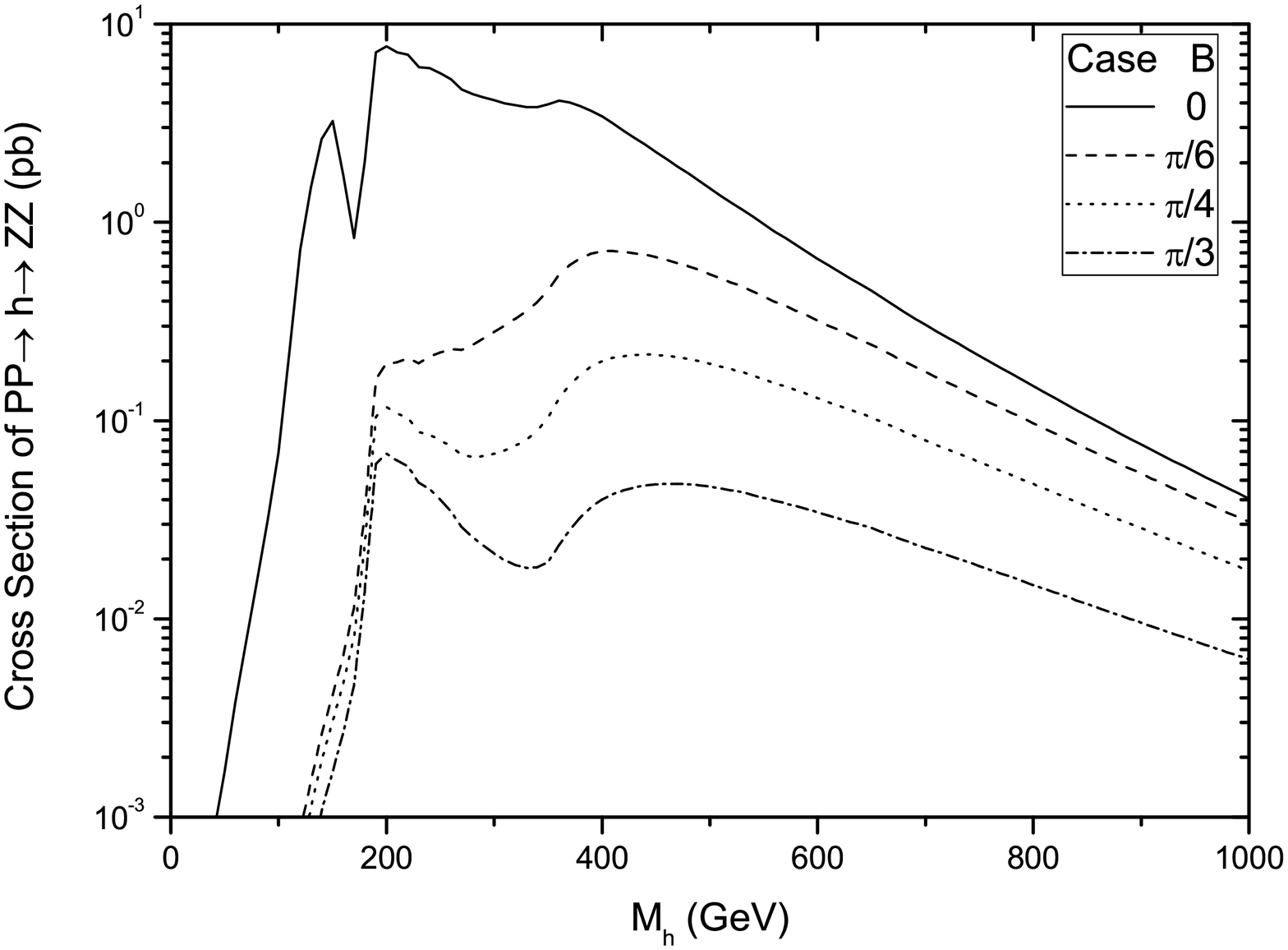}\\
  \includegraphics[scale=0.35]{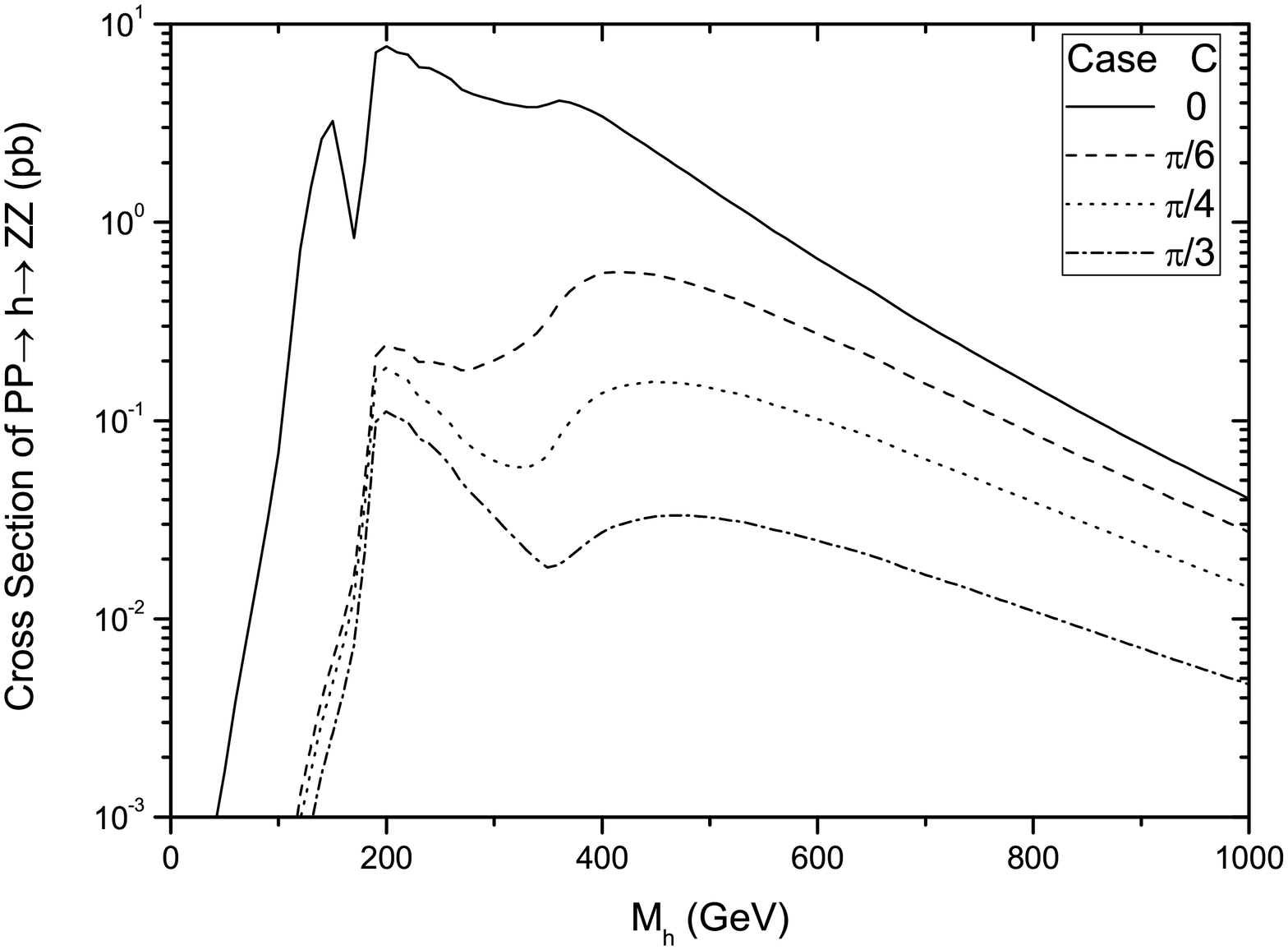}
\caption{The Higgs mixing effect to the cross section of $PP\to h\to
ZZ$ which is regarded as golden channel to search for a heavy
SM-like Higgs $h$. The solid line is for the case without mixing
which is similar to SM, while other lines are for the cases with
mixing angles: $\theta=\pi/6, \pi/4, \pi/3$.}
  \label{fig:mixing_h1zz}
\end{figure}

\begin{figure}[htb]
  \includegraphics[scale=0.35]{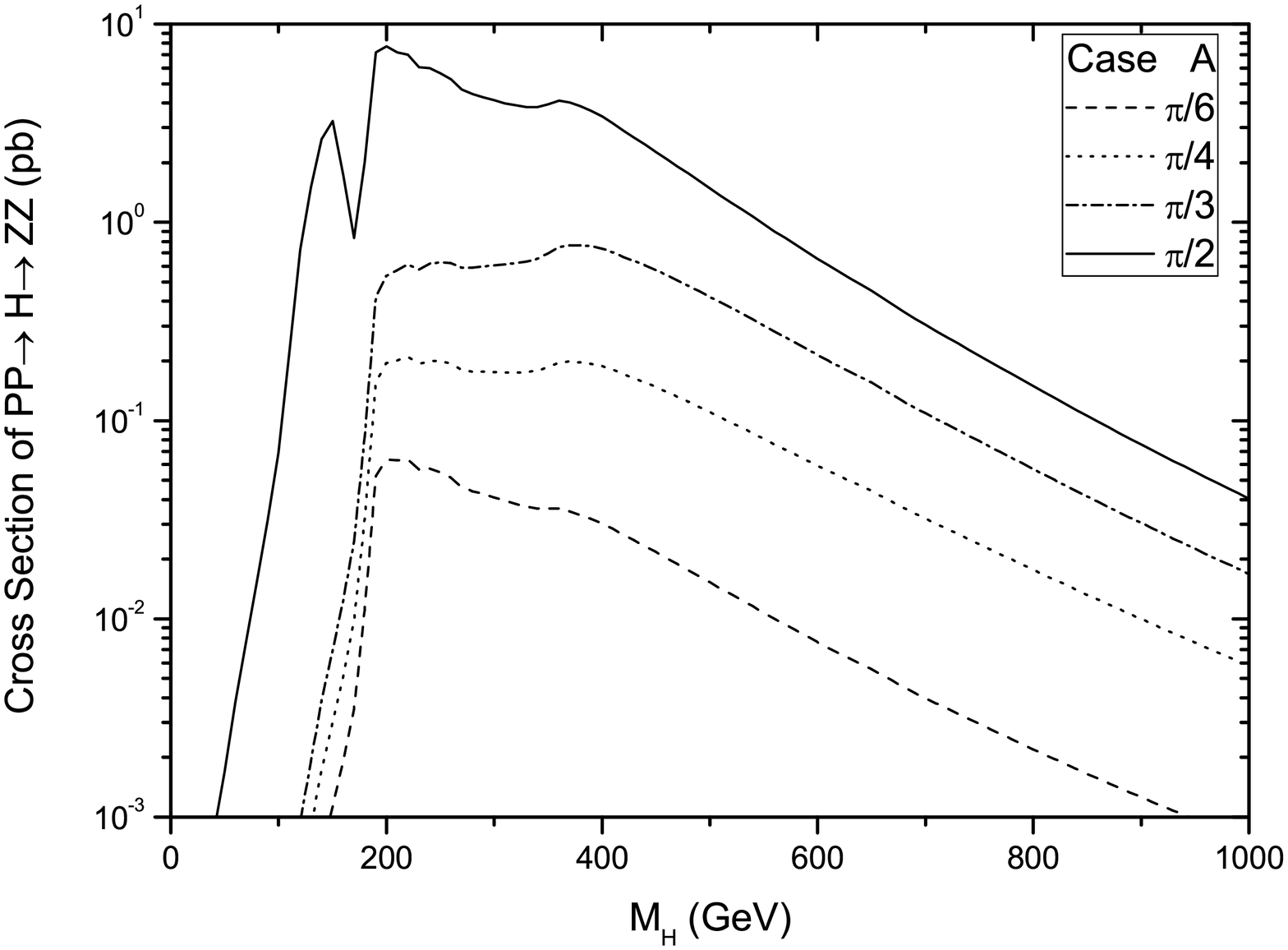}
  \includegraphics[scale=0.35]{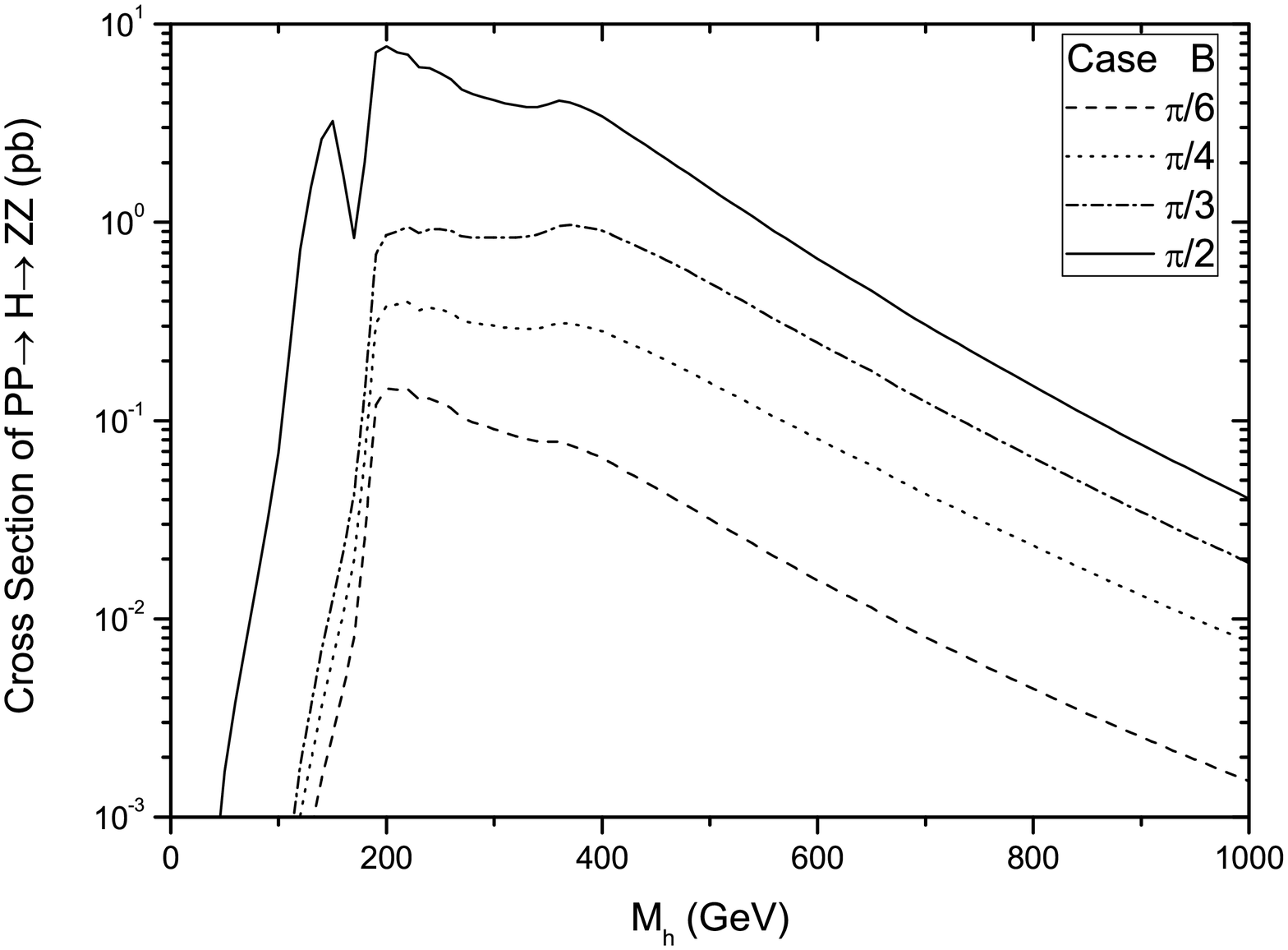}
  \includegraphics[scale=0.35]{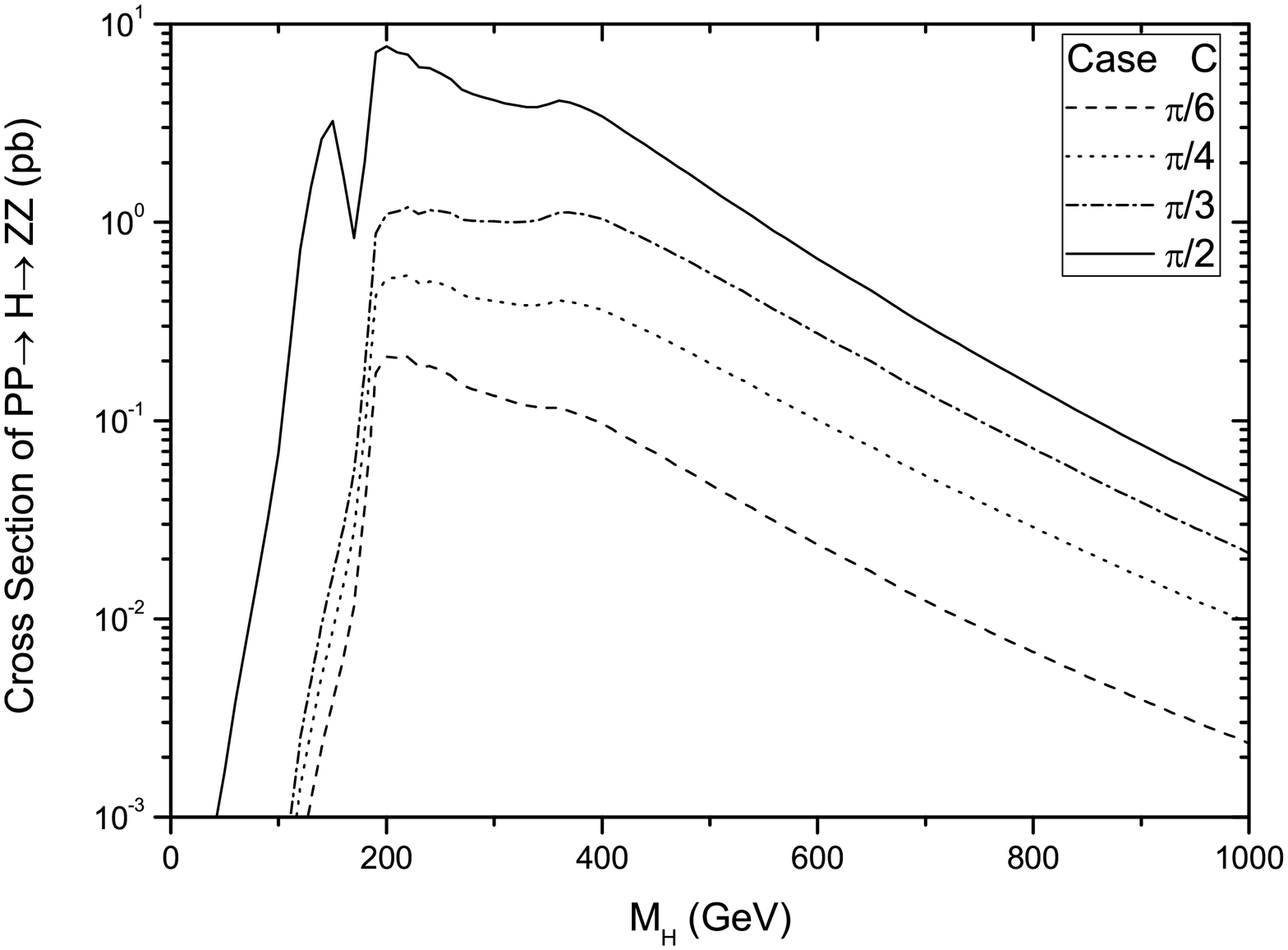}
\caption{The Higgs mixing effect to the cross section of $PP\to h\to
ZZ$. For $\theta=0$, the $H$ can not decay to ZZ at tree level, but
for $\theta=\pi/2$, the $H$ plays the role of the Higgs in SM as
shown from the solid lines.}
  \label{fig:mixing_h2zz}
\end{figure}

\begin{figure}[htb]
\includegraphics[scale=0.35]{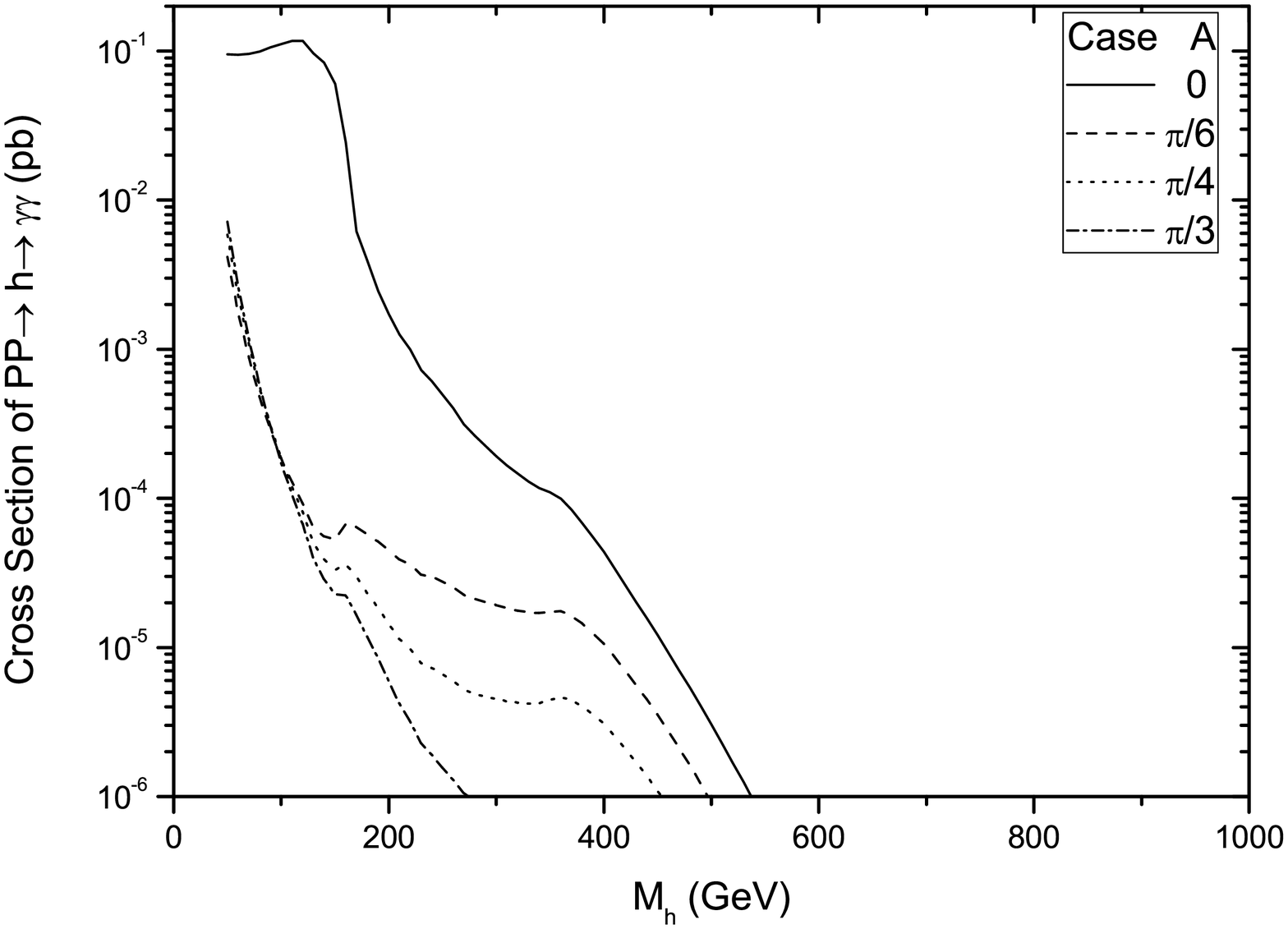}
\includegraphics[scale=0.35]{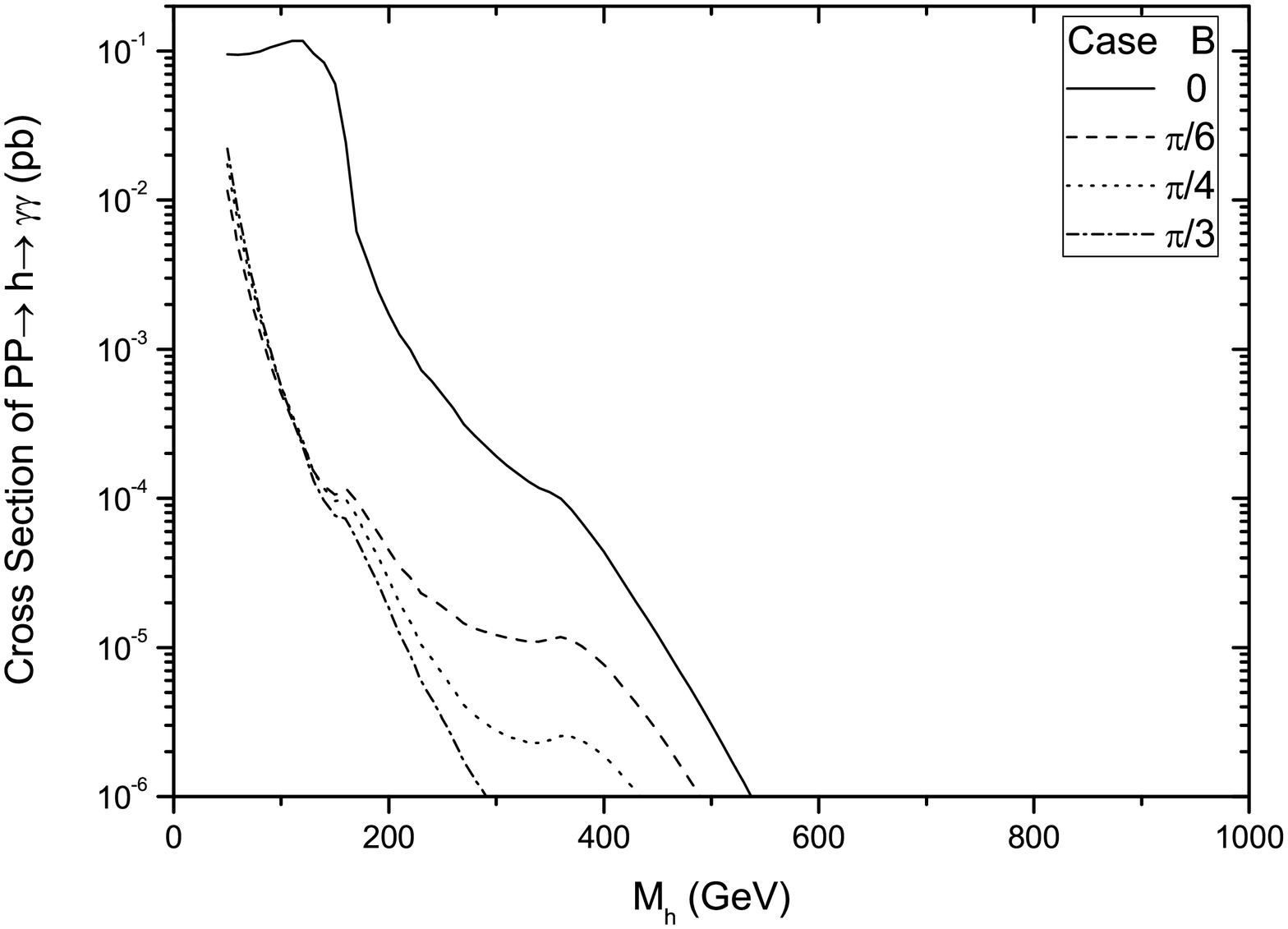}
\includegraphics[scale=0.35]{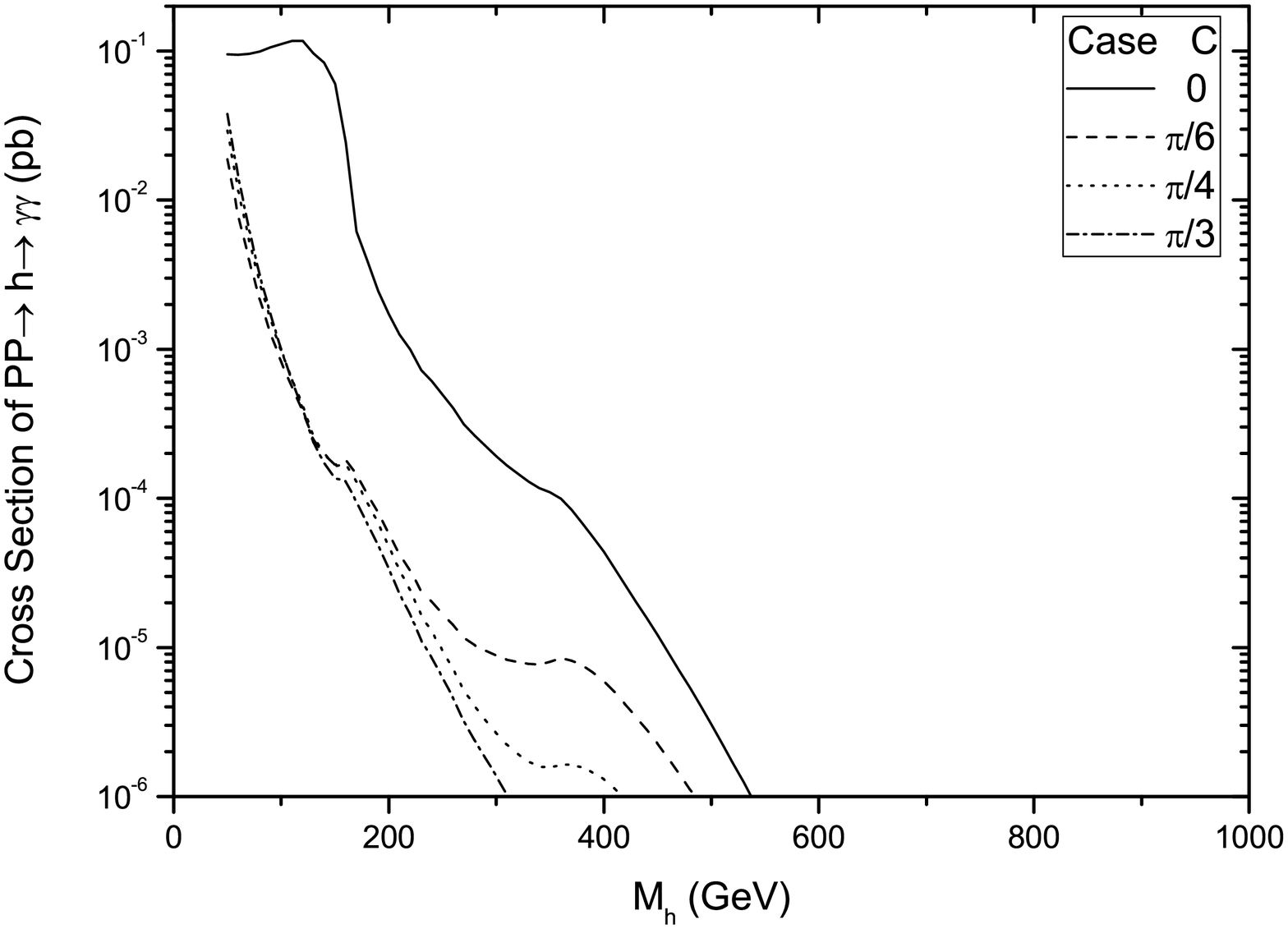}
\caption{The Higgs mixing effect to the cross section of $PP\to h\to
\gamma\gamma$ which is regarded as golden channel to search for a
heavy SM-like Higgs $h$. The solid line is for the case without
mixing which is similar to SM, while other lines are for the cases
with mixing angles: $\theta=\pi/6, \pi/4, \pi/3$.}
\label{fig:mixing_gm1}
\end{figure}

\begin{figure}[htb]
\includegraphics[scale=0.35]{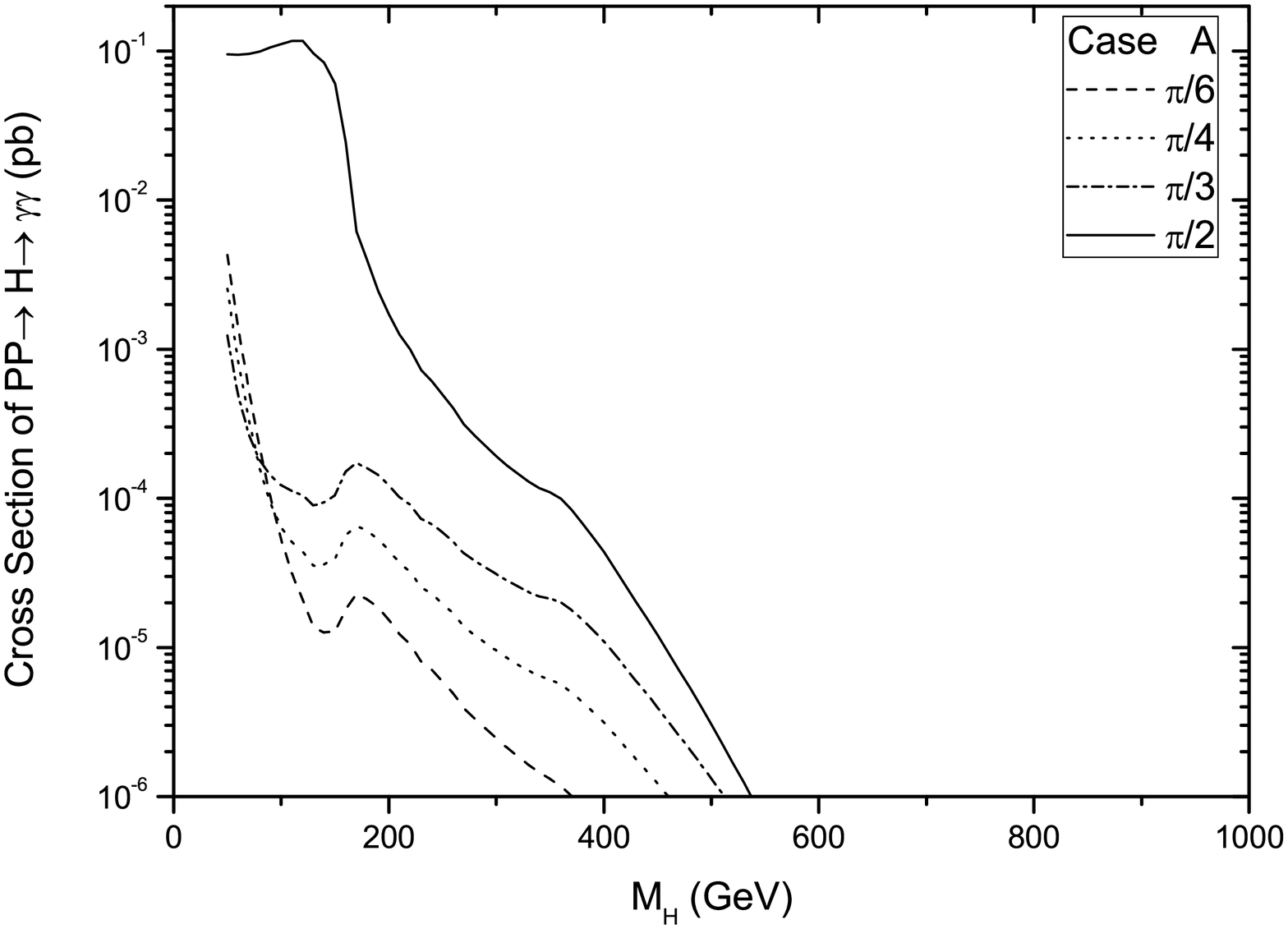}
\includegraphics[scale=0.35]{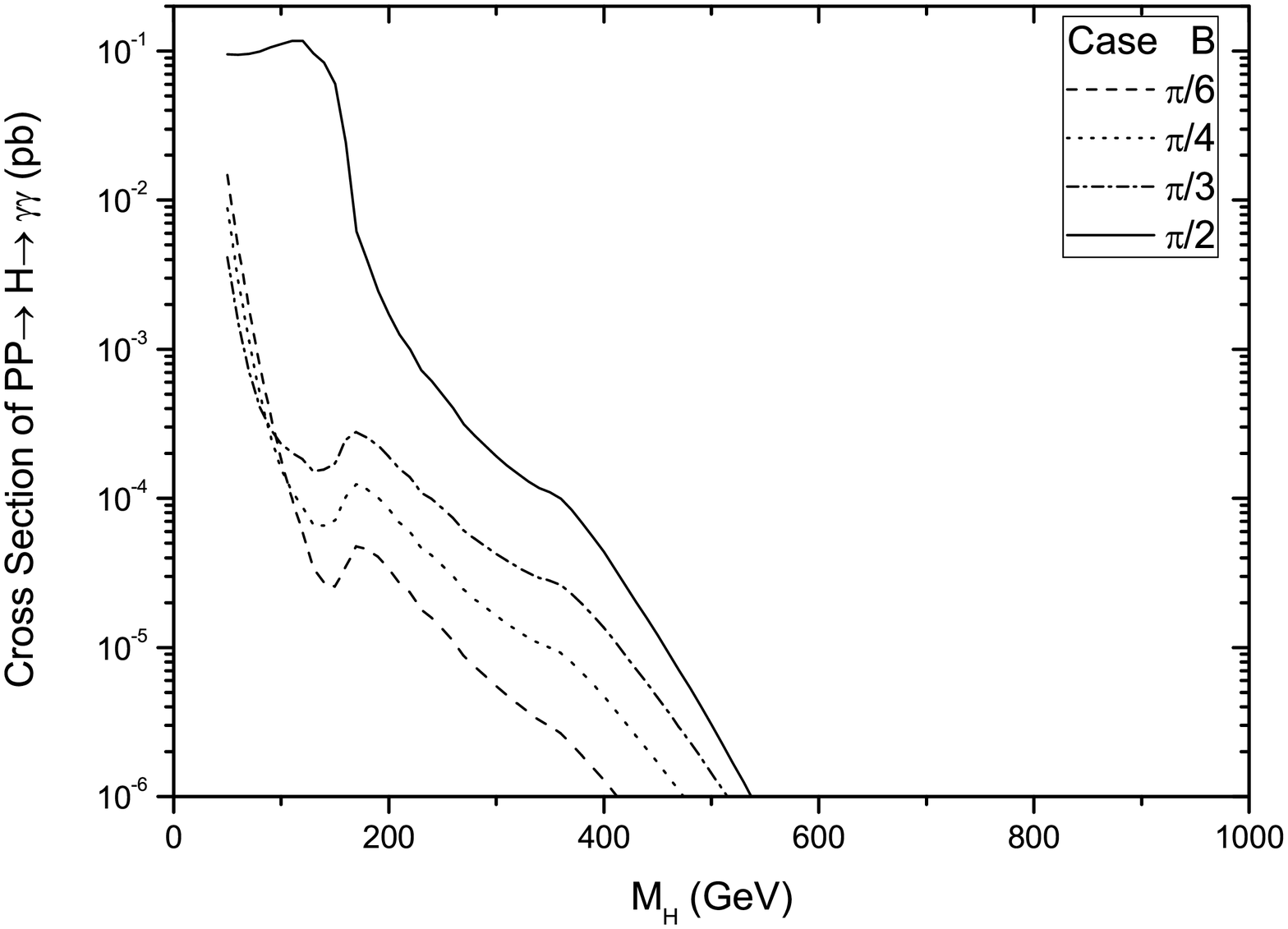}
\includegraphics[scale=0.35]{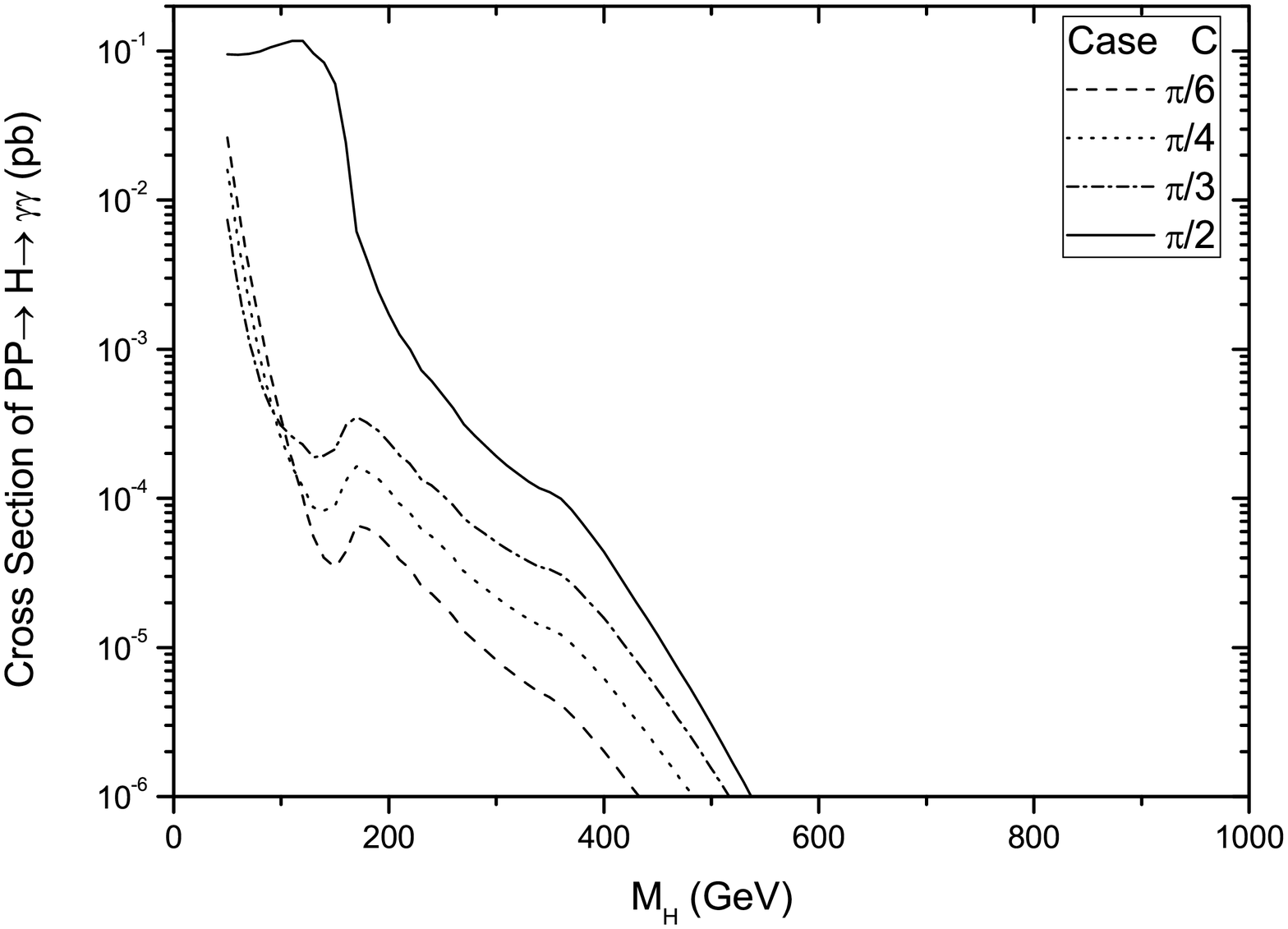}
\caption{The Higgs mixing effect to the cross section of $PP\to h\to
\gamma\gamma$. For $\theta=\pi/2$, the $H$ plays the role of the
Higgs in SM as shown from the solid lines.} \label{fig:mixing_gm2}
\end{figure}

Our results are shown in Fig.\ref{fig:mixing_h1zz}, Fig.\ref{fig:mixing_h2zz},
Fig.\ref{fig:mixing_gm1} and Fig.\ref{fig:mixing_gm2}. From the
Fig.\ref{fig:mixing_h1zz} and Fig.\ref{fig:mixing_gm1}, it can be
seen that the mixing effects could become significant for the
SM-like Higgs $h$, its cross section can be much suppressed when the
mixing angle becomes large.

\section{conclusion\label{section:conclusion}}

We have investigated, in the general 2HDM with spontaneous $CP$
violation, the production and decays of the SM-like Higgs $h$ and
the new neutral Higgs $H$. Numerically, we have considered three
typical sets of Yukawa couplings for the Higgs boson $H$, which is
consistent with the current experimental bounds from the flavor
sector even when the Higgs boson mass is as low as $M \simeq 160$
GeV. It has been seen that when $h$ and $H$ are both light, i.e., $M
< 200$ GeV, the production cross section of $H$ is in general larger
than the one of $h$, while for $M > 200$ GeV, the production cross
section of $h$ becomes larger than the one of $H$. As the Yukawa
couplings of $H$ can be complex, its production cross section can
strongly rely on the $CP$-violating phase and be affected
significantly. The $b\bar{b}$ decay mode is the dominant channel for
both $h$ and $H$, while the $H\to \gamma\gamma$ and $H\to gg$ are
both smaller than $h\to \gamma\gamma$ and $h\to gg$, thus it is not
difficult to distinguish them. The SM-like Higgs $h$ can be detected
via the golden channel $h\to ZZ\to 4l$, while the new neutral Higgs
$H$ has no such a channel at tree level if without considering the
neutral Higgs mixing, it is mainly detected via $H\to \bar{b}b$ as
it can be very different from $h\to \bar{b}b$ due to different
Yukawa coupling. When the mixing between $h$ and $H$ becomes very
large and their mass difference is very small, it is then not very
easy to distinguish them from the production signals. It is noted
that LHC does not favor 2HDM with all parameter spaces, especially
in the decoupling limit with a small Yukawa coupling $\xi^f \ll
m_f/v$ and a small mixing $\theta\simeq 0$ between $h$ and $H$, in
this case $h$ looks the same as the SM Higgs and it then becomes
hard to detect the new Higgs bosons, thus one is not able to
distinguish the Type III 2HDM and SM from a direct detection. In
general, the mixing between the neutral Higgs bosons $h$ and $H$ is
characterized by a free parameter $\theta$ which can be large, so
that the production cross section and decays of the neutral Higgs
boson can significantly be different from the predictions from the
standard model. It would be very interesting to search for the
possible new Higgs boson effects at LHC or at ILC.

\begin{acknowledgments}
The authors would like to thank W.Y. Wang and Y. L. Ma for useful
discussions. This work was supported in part by the National Science
Foundation of China (NSFC) under the grant \# 10821504 and the key
Project of Knowledge Innovation Program (PKIP) of Chinese Academy of
Science.
\end{acknowledgments}

\newpage

\begin{thebibliography}{99}
\bibitem{unitarity}
B. W. Lee, C. Quigg and H. B. Thacker, \prl{38}, 883 (1977);
\prd{16}, 1519 (1977).
\bibitem{lep}
D. Abbaneo \etal  (LEP Electroweak Working Group) and (SLD
Electroweak and Heavy Flavor Groups), arXiv: hep-ex/0312023;  R.
Barate \etal Phys. Lett. B 565, 61 (2003).
\bibitem{higgsarg}
T.Hambye and K. Riesselmann, \prd {55}, 7255 (1997); M.~Lindner, Z.\
Phys.\ C {\bf 31}, 295 (1986).
\bibitem{TDLEE}
T. D. Lee, Phys. Rev. D {\bf 8}, 1226 (1973); Phys. Rep. {\bf 9},
143 (1974).
\bibitem{SW}
S. Weinberg, Phys. Rev. Lett. {\bf 37}, 657(1976).

\bibitem{YLW1}
Y.L. Wu, Carnegie-Mellon Univ. report, CMU-HEP94-01, hep-ph/9404241,
1994; in {\it Proceedings at 5th Conference on the Intersections of
Particle and Nuclear Physics, St. Petersburg, FL, 31 May- 6 Jun
1994}, edited by S. J. Seestrom (AIP, New York, 1995), p.338.
\bibitem{YLW2}
Y. L. Wu and L. Wolfenstein, Phys. Rev. Lett 73, 1762 (1994).
\bibitem{cpv}
G. C. Branco, L. Lavoura and J. P. Silva, {\it CP Violation},
(Oxford University Press, Oxford, 1999).
\bibitem{KM}
  M.~Kobayashi and T.~Maskawa, Prog.\ Theor.\ Phys.\  {\bf 49}, 652 (1973).
\bibitem{PR}D. Atwood, L. Reina and A. Soni, Phys. Rev. D 55
3156 (1997); D. Atwood, S. Bar-Shalom, G. Eilam, A. Soni, Phys.
Rept. {\bf347}, 1 (2001), and reference therein.
\bibitem{discrete}S. Glashow and S. Weinberg, \prd{\bf15}, 1958
(1977).
\bibitem{barger1} V.~D.~Barger, J.~L.~Hewett and R.~J.~N.~Phillips,
  Phys.\ Rev.\  D {\bf 41}, 3421 (1990).
\bibitem{barger2}  V.D. Barger, H. E. Logan and G. Shaughnessy, Phys.\ Rev.\ D {\bf 79}, 115018 (2009).
\bibitem{noscpv}G. C. Branco, and M. N. Rebelo, Phys. Lett. B {\bf160},
117(1985).
\bibitem{YLW3} L. Wolfenstein and Y.L. Wu, Phys. Rev. Lett. {\bf 73} 2809
(1994).
\bibitem{FS}A. Antaramian, L. J. Hall, and A. Rasin, Phys. Rev. Lett. {\bf69},1871 (1992);
M. J. Savage, Phys. Lett. {\bf B266}, 135 (1991); W. S. Hou, Phys.
Lett. {\bf B296}, 179 (1992); D. W. Chang, W. S. Hou and W. Y.
Keung, Phys. Rev. D {\bf 48}, 217(1993);
\bibitem{HW} L.J. Hall and S. Weinberg, Phys.\ Rev.\  D {\bf 48}, R979 (1993)
\bibitem{BCK}
D. Bowser-Chao, K. Cheung, and W.-Y. Keung, Phys. Rev. D {\bf 59},
115006, (1999).
\bibitem{WUZHOU} Y.L. Wu and Y.F. Zhou, Phys. Rev. D {\bf 61} 096001
(2000).
\bibitem{2hdmeff}
C. S. Huang and S. H. Zhu, \prd{68}, 114020(2003); Y. B. Dai, C. S.
Huang,J. T. Li and W. J. Li, Phys. Rev. D {\bf 67}, 096007 (2003).
\bibitem{YLW4} Y.L. Wu, Chin. Phys. Lett. {\bf16} 339 (1999).
\bibitem{b2gamma}
  C. S. Huang and J. T. Li,
Int. J. Mod. Phys. A 20, 161 (2005).
\bibitem{B2PV}
Y. L. Wu and C. Zhuang, Phys. Rev. D {\bf 75}, 115006 (2007).
\bibitem{B2VV}
  S.~S.~Bao, F.~Su, Y.~L.~Wu and C.~Zhuang, Phys.\ Rev.\  D {\bf 77}, 095004 (2008).
\bibitem{paire_pro}A.~Arhrib, R.~Benbrik, C.~H.~Chen, R.~Guedes and R.~Santos,
  JHEP {\bf 0908}, 035 (2009).
\bibitem{Han:2005mu}
  T.~Han, in {\it Physics in D$>=4$, Proceedings of the Theoretical Advance Study Institute in Elementary Particle Physics, Boulder, Colorado, 6 Jun 2 Jul 2004,} edited by J. Terning, C. E. M. Wagner, and D. Zeppenfield (World Scientific, Singapore, 2006), p. 407.
\bibitem{collider}
Vernon D. Barger and Roger J. N. Phillips, {\it Collider Physics},
(Addison-Wesley, Redwood City, USA, 1987).
\bibitem{georgi}
H. M. Georgi, S. L. Glashow, M. E. Machacek and D. V. Nanopoulos, Phys. Rev. Lett, {\bf 40} 692(1978).

\bibitem{Graudenz}
D. Graudenz, M. Spira, P. M. Zerwas. Phys. Lett. {\bf 70}, 1372(1993).
\bibitem{Marzani:2008az}
  S.~Marzani, R.~D.~Ball, V.~Del Duca, S.~Forte and A.~Vicini,
  Nucl.\ Phys.\  B {\bf 800}, 127 (2008).
\bibitem{hproduction}
  M.~Spira, A.~Djouadi, D.~Graudenz and P.~M.~Zerwas,
  Nucl.\ Phys.\  B {\bf 453}, 17 (1995).
\bibitem{NNLO}
  R.~V.~Harlander and W.~B.~Kilgore,
  Phys.\ Rev.\ Lett.\  {\bf 88}, 201801 (2002).
\bibitem{NNLO1}
  C.~Anastasiou and K.~Melnikov,
  Nucl.\ Phys.\  B {\bf 646}, 220 (2002);
  V.~Ravindran, J.~Smith and W.~L.~van Neerven,
  Nucl.\ Phys.\  B {\bf 665}, 325 (2003).
\bibitem{NNNLO}
S. Moch and A. Vogt, Phys. Lett. B {\bf 631}, 48(2005); E. Laenen
and L. Magnea,  Phys. Lett. B {\bf 631}, 48(2005).
\bibitem{NNLO2}
A. Idilbi \etal, Phys. Rev. D {\bf 73} 077501(2006); V. Ravindran,
Nucl. Phys. B {\bf 752}, 173(2006).
\bibitem{EWC}
R.~Bonciani,  J.\ Phys.\ Conf.\ Ser.\  {\bf 110}, 042004 (2008); A.
Djouadi and P. Gambino, Phys. Rev. Lett. {\bf 73} 2528(1994);
 U.~Aglietti, R.~Bonciani, G.~Degrassi and A.~Vicini,
  Phys.\ Lett.\  B {\bf 595}, 432 (2004).
\bibitem{pdg}
 Particle Data Group, C. Amsler \etal., Physics Letters {\bf B667}, 1
 (2008).
\bibitem{cteq}
  H.~L.~Lai {\it et al.}  [CTEQ Collaboration],
  Eur.\ Phys.\ J.\  C {\bf 12}, 375 (2000).
\bibitem{gamma5}
S.L. Adler and W.A. Bardeen, Phys. Rev. 182 1517 (1969);
 W. Bardeen, Phys. Rev. 184 1848 (1969).
\bibitem{wugamma5}
  Y.~L.~Ma and Y.~L.~Wu, Int.\ J.\ Mod.\ Phys.\  A {\bf 21}, 6383 (2006).
\bibitem{hdecay}
  A.~Djouadi, J.~Kalinowski and M.~Spira,
  Comput.\ Phys.\ Commun.\  {\bf 108}, 56 (1998).
\bibitem{ggtt}G.~Aad {\it et al.}  (The ATLAS Collaboration),
  arXiv:0901.0512;
    G.~L.~Bayatian {\it et al.}  (CMS Collaboration),
  J.\ Phys.\ G {\bf 34}, 995 (2007).
\bibitem{linear}
  J.~A.~Aguilar-Saavedra {\it et al.}  (ECFA/DESY LC Physics Working Group),
  arXiv:hep-ph/0106315;
    K.~Abe {\it et al.}  (ACFA Linear Collider Working Group),
  arXiv:hep-ph/0109166;
 E.~Accomando {\it et al.} (CLIC Physics Working Group),
  arXiv:hep-ph/0412251;
   G.~Weiglein {\it et al.}  (LHC/LC Study Group),
  Phys.\ Rept.\  {\bf 426}, 47 (2006).
\bibitem{linear2} D.~Lopez-Val and J.~Sola,
  Phys.\ Rev.\  D {\bf 81}, 033003 (2010);
   R.~N.~Hodgkinson, D.~Lopez-Val and J.~Sola,
  Phys.\ Lett.\  B {\bf 673}, 47 (2009);
   E.~Asakawa, D.~Harada, S.~Kanemura, Y.~Okada and K.~Tsumura,
  arXiv:0902.2458;
  Phys.\ Lett.\  B {\bf 672}, 354 (2009);
  A.~Arhrib, R.~Benbrik, C.~H.~Chen and R.~Santos,
  Phys.\ Rev.\  D {\bf 80}, 015010 (2009).
\bibitem{associated}
A.~Del Fabbro and D.~Treleani,
  Phys.\ Rev.\  D {\bf 61}, 077502 (2000);
A. De Roeck and G. Polesello, C. R. Physique 8, 1078 (2007); J. M.
Butterworth, A. R. Davison, M. Rubin and G. P. Salam, Phys. Rev.
Lett. 100, 242001 (2008).
\bibitem{gammagamma} D.~d'Enterria and J.~P.~Lansberg,
  Phys.\ Rev.\  D {\bf 81}, 014004 (2010).
\end{thebibliography}
\end{document}